\documentclass[twocolumn,10pt,aps,pra,superscriptaddress,nofootinbib]{revtex4-1}

\usepackage{scrextend}
\usepackage{extra_style}
\usepackage{longtable}
\usepackage[a4paper,bindingoffset=0.2in,left=.75in,right=.75in,
            top=1.5in,
            bottom=1.25in,
            footskip=.005in]{geometry}
\usepackage{graphicx}
\renewcommand{\thefootnote}{$\star$} 
\renewcommand*{\thefootnote}{\fnsymbol{footnote}}

\newcommand{\SuppI}{Appendix}

\usepackage{xspace}

\usepackage{silence}
\newcommand{\xhdr}[1]{\vspace{1.7mm}\noindent{{\bf #1.}}}

\begin{document}

\title{Meat-Free Day Reduces Greenhouse Gas Emissions but Poses Challenges for Customer Retention and Adherence to Dietary Guidelines}
\author{ Giuseppe Russo,$^*$ 
Kristina Gligorić,$^\dagger$
Vincent Moreau,$^*$\\
Robert West$^*$\\ 
{$^*$ EPFL, $^\dagger$ Stanford University} \\
}

\begin{abstract}
\noindent 
Reducing meat consumption is crucial for achieving global environmental and nutritional targets. Meat-Free Day (MFD) is a widely adopted strategy to address this challenge by encouraging plant-based diets through the removal of animal-based meals. We assessed environmental, behavioral, and nutritional impacts of MFD by implementing 67 MFDs over 18 months (once a week on a randomly chosen day) across 12 cafeterias on a large university campus, analyzing over 400{,}000 food purchases. MFD reduced on-campus food-related greenhouse gas (GHG) emissions on treated days by 52.9\% and contributed to improving fiber (+26.9\%) and cholesterol ($-$4.5\%) consumption without altering caloric intake. These nutritional benefits were, however, accompanied by a 27.6\% decrease in protein intake and a 34.2\% increase in sugar consumption. 
Moreover,
the increase in plant-based meals did not carry over to subsequent days, as evidenced by a 3.5\% rebound in animal-based meal consumption on days immediately following treated days,
and
MFD led to a 16.8\% drop in on-campus meal sales on treated days.
Monte Carlo simulations suggest
that, if 8.7\% of diners were to eat burgers off-campus on treated days, MFD’s GHG savings would be fully negated.
As our analysis identifies on-campus customer retention as the main challenge to MFD effectiveness, we recommend combining MFD with customer retention interventions to ensure  environmental and nutritional benefits.
\end{abstract}

\maketitle

\footnotetext{Correspondence:\ 
{giuseppe.russo@epfl.ch},
{robert.west@epfl.ch}
}
\renewcommand*{\thefootnote}{\arabic{footnote}}
\setcounter{footnote}{0}

\noindent Current consumption of animal-derived food is incompatible with the sustainability and nutritional targets set by the UN Sustainability and Nutrition Strategy 2022-2030 \cite{un2022nutrition, un2023sdg, springmann2018options}. Achieving these goals requires wealthy countries to reduce their meat consumption by 80\%,  a weekly decrease of 2.4 kg/person of meat eaten \cite{willett2019food, oecd2021meat, parlasca2022meat}.
A progressive transition to plant-based diets emerges as the most effective strategy to (i) mitigate the environmental impact of the global food systems \cite{earthday2019, frontiers2021, tilman2014} and (ii) provide nutritional benefits \cite{sciencedirect2019, academy2016}. 
Appropriately planned and supplemented vegan and vegetarian diets are nutritionally adequate for individuals in all stages of life and may provide health benefits in disease prevention \cite{wang2023vegetarian, kahleova2018vegetarian}. In contrast, the consumption of animal-based meals is a core driver of large shares of carbon emissions~\cite{henry2019role, xu2021global} and has been associated with between 3\% to 7\% higher risk of cardiovascular disease and mortality from all causes \cite{zhong2020associations, russo2024stranger, micha2010red}.

Shifting behavior towards plant-based diets is difficult due to key societal barriers, including misconceptions about the nutritional value of plant-based options \cite{feher2020comprehensive, giacalone2022understanding}, higher perceived costs of plant-based foods \cite{estell2021plant, varela2022meat}, socioeconomic differences \cite{klink2022socioeconomic, schuz2021equity} and cultural norms \cite{rothgerber2013real}.

In response to these barriers, governments and institutions have funded various initiatives aimed at reducing meat consumption.
The most prominent of these initiatives is ``Meat-Free Day'' (MFD). 
Proposed more than 20 years ago~\cite{meatless_monday_origin, veggie_day_ghent}, MFD aims to reduce the environmental impact of food consumption and promote healthy nutrition with one simple policy: ``Once a week, cut meat.'' Ever since, MFD has become a global movement implemented in over 40 countries \cite{time_meatless_monday, nytimes_meatless_monday, meat_free_movement},  hospitals \cite{forbes_meatless_hospitals}, schools \cite{lombardini2013forced}, and university campuses \cite{harvard_meatless}, affecting more than $10,000$ cafeterias and millions of people. 
MFD has been a controversial initiative that has also sparked political discussion~\cite{politico_mfd_political, dw_veggie_day_debate}, polarized public opinion~\cite{nyt_meatout_2021}, and fueled debates across multiple legislative bodies~\cite{congress_meatless_monday, german_parliament_meatless_day}, such as the US Congress and German Parliament.
Proponents praise it for potential health and environmental benefits, whereas opponents criticize it as an ineffective restriction on personal choice that is unlikely to push people towards sustainable habits.

Despite the widespread interest in the MFD initiative, existing studies have only partially addressed its effectiveness~\cite{lambrecht2023limiting, de2012nutritional, deboer2014attitudes}. Most analyses have focused on changes in environmental outcomes, such as greenhouse gas (GHG) emissions or water footprint \cite{lambrecht2023limiting, blondin2022meatless}, overlooking other critical dimensions including nutritional impacts and subsequent consumption of plant-based meals. Furthermore, the estimation of meal-related emissions has often relied on high-level proxy measures due to the absence of detailed, ingredient-level data \cite{merk2024no}. Critically, these studies rarely account for unintended behavioral responses~\cite{lombardini2013forced, lambrecht2023limiting}, which are essential in assessing the effectiveness of any policiy \cite{russo2023spillover, latona2024ai}. Even when compensatory behaviors are considered, evaluations typically occur under unrealistic conditions—such as infrequent scheduling \cite{merk2024no} or scenarios lacking alternative animal-based meal options \cite{lombardini2013forced}—severely limiting the assessment of MFD effectiveness.

In this work, we overcome these limitations by conducting a large-scale evaluation of the MFD initiative. Our analysis leverages detailed transactional data alongside precise ingredient-level emissions and nutritional information to rigorously assess the primary objectives of the MFD initiative: (i) institutional reductions in GHG emissions, (ii) behavioral shifts toward reduced meat consumption, and (iii) enhancements in nutritional quality aligned with established dietary guidelines.

\begin{figure*}[t]
    \centering
    \includegraphics[width=1.0\textwidth]{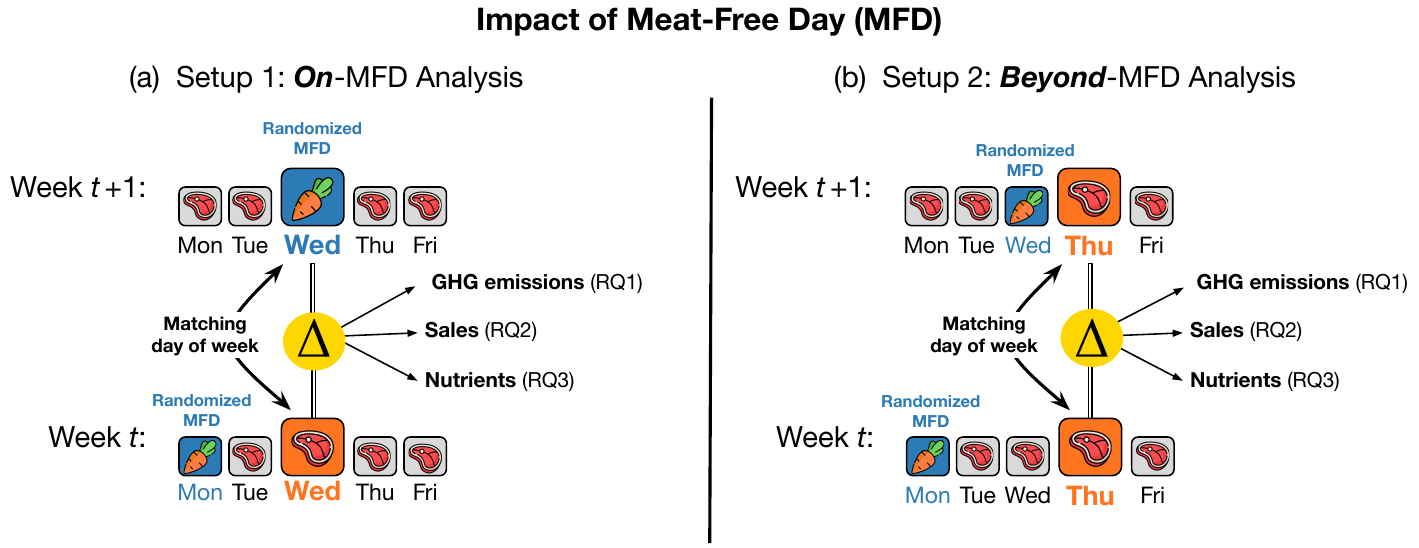}

\caption{\footnotesize \justifying \textbf{Analysis setups to estimate the effect of Meat-Free Day (MFD).} 
\textbf{(a) on-MFD Analysis:} This setup examines the direct impact of MFD on three key outcomes: greenhouse gas (GHG) emissions (\textbf{RQ1}), sales patterns (\textbf{RQ2}), and nutrient intake (\textbf{RQ3}). Each MFD (randomly assigned to a weekday in week \( t+1 \)) is matched with a corresponding non-MFD from the preceding week \( t \), occurring on the same weekday. This comparison quantifies the immediate effect \( \Delta \) on these outcomes.
\textbf{(b) beyond-MFD Analysis:} This setup investigates potential rebound effects by analyzing consumer behavior on the day following an MFD. A non-MFD in week \( t+1 \) that occurs immediately after week \(t+1\)'s MFD is matched with a non-MFD from week \( t \) that occurs on the same weekday, but at least two days after week \(t\)'s MFD. This comparison assesses whether MFD triggers compensatory behaviors that could counteract its intended effects.
}

    \label{fig:overview}
    \vspace{-1mm}

\end{figure*}

\xhdr{Present work} For each of these objectives, we formulate a corresponding research question:

\textbf{RQ1:} Does restricting the availability of animal-based meals via MFD lead to a measurable reduction in the GHG emissions associated with food purchases? 

\textbf{RQ2:} Does the  restriction of choice via MFD trigger behavioral responses that counteract its intended effects?

\textbf{RQ3:} What are the nutritional implications of MFD, and does the removal of meat-based meals lead to compensatory consumption of alternatives?

To answer these research questions, we implemented the Meat-Free Day initiative 67 times across all 12 cafeterias on the EPFL (Ecole Polytechnique Fédérale de Lausanne) campus in Lausanne, Switzerland. The intervention was conducted once a week over 18 months, with a single day each week randomly selected to serve as an MFD. This random selection was made by the administration without prior notice to the campus community. In total, to study the impact of these 67 MFDs, we analyzed $402{,}536$ food purchases. 

In \emph{RQ1}, we assess the impact of MFD on GHG emissions associated with on-campus food consumption. By analyzing purchasing patterns across university cafeterias, we quantify how the removal of animal-based meals directly influences the GHG emissions of the campus food system.

In \emph{RQ2}, we test the hypothesis that perceived restriction of choice leads to behavioral responses that could counteract MFD’s intended effects~\cite{brehm1966theory, vandenbroele2020nudging}. Specifically, we examine two mechanisms: (i) \emph{opt-out behavior}, where customers circumvent MFD by eating off-campus, presumably consuming animal-based meals, and (ii) \emph{rebound effects}, where individuals increase their consumption of animal-based meals on subsequent days to compensate.

In \emph{RQ3}, we assess the nutritional consequences of MFD by analyzing shifts in calories and macronutrient intake. While plant-based diets provide health benefits \cite{craig2021safe, hemler2019plant}, the removal of animal-based meals may introduce unintended dietary imbalances. Additionally, previous research indicates that customers often perceive vegetarian meals as less nutritious \cite{appleton2016barriers, jahn2021plant, kemper2020motivations}, which may prompt them to compensate by purchasing energy-dense snacks or drinks from vending machines and cafeterias \cite{gligoric2024measuring}.

To evaluate the effects of MFD, we adopt two complementary setups. The first setup, which we term \emph{on-MFD Analysis}, examines the immediate impact of MFD on the day of intervention, assessing its effects on three outcomes: (i)  GHG emissions ({RQ1}), (ii) opt-out behavior ({RQ2}), and (iii) caloric and macronutrient intake ({RQ3}).
To measure these changes, we compared each MFD with a corresponding non-MFD on the same weekday of the preceding week (see Fig. \ref{fig:overview}(a)).

The second setup, which we term \emph{beyond-MFD Analysis}, extends the investigation to the days following MFD, evaluating whether the intervention produces beneficial changes in customers' eating habits or triggers unintended compensatory behaviors that could reduce the intended benefits. To assess this, we analyze variations in the three outcomes described for the \emph{on-MFD Analysis} on two non-MFDs from consecutive weeks: one occurring the day after an MFD and another occurring at least two days after the MFD of its respective week (see Fig. \ref{fig:overview}(b)).

\xhdr{Findings} Our findings demonstrate that MFD-treated days achieve a remarkable 52.9\% reduction in food-system GHG emissions compared to untreated days.
 (\emph{RQ1, on-MFD Analysis}).
Although this reduction underscores the environmental benefits of MFD, we also identify unintended consequences: meal sales drop by 16.8\%, which translates to 520 fewer meals sold compared to the daily baseline of 3{,}092 meals on non-MFDs (\emph{RQ2, on-MFD Analysis}).
This decline suggests a behavioral shift, likely driven by the perceived restriction of choice, with customers opting out of MFD to eat off-campus.

Opt-out customers are likely to consume animal-based meals off-campus. They may choose from either a nearby fast-food restaurant serving burgers or other off-campus options offering animal-based meals with lower emissions than burgers. Due to the lack of data on off-campus dining sales, we used Monte Carlo simulations to estimate the impact of off-campus dining on MFD’s overall GHG emission savings. This analysis revealed that, even assuming that all opt-out customers choose the lower-emission options, 58.6\% of MFD’s GHG emission savings could be offset by these choices. Our simulation indicates that the GHG emission savings achieved by MFD would be fully offset by off campus diners if 50.1\% of them ate burgers. That corresponds to 8.7\% of the campus community that dine in cafeterias on a typical day.

Furthermore, we found no evidence that exposure to vegetarian meals during MFD fosters plant-based eating habits beyond the intervention day. 
We observe a statistically significant increase of 4.3\% in GHG emissions on the day following an MFD, 
with a 3.5\% rise in animal-based meal sales compared to non-MFDs occurring at least two days after an MFD (\emph{RQ1 and RQ2, beyond-MFD Analysis})).

Regarding the impact of MFD on nutritional value, we find no measurable change in total caloric intake, suggesting that implementing MFD does not inherently reduce overall calories. 

However, the impact on macronutrient intake is mixed. An increase in fiber (+27.2\%) and the reduction in cholesterol ($-4.5\%$) suggest potential health benefits. At the same time, an increase in sugar intake (+39.9\%) and a decrease in protein intake ($-27.1\%$) may warrant attention if sustained over time. While there is no  increase in the consumption of unhealthy items (e.g., desserts, vending machine items) overall, we found a 15.9\% increase in dessert purchases, maybe to offset a perceived decrease in calories~\cite{erfanian2024cultivating}---a common bias observed in customers consuming plant-based meals. 
However, we find no measurable change in total caloric,  macronutrient, and unhealthy-item intake on days immediately following MFD-treated days.

\xhdr{Implications} Our work shows that the Meat-Free Day initiative is not a silver bullet.
From a nutritional perspective, while MFD can introduce positive dietary shifts such as increased fiber consumption, the concurrent increase in sugar intake and reduction in protein highlights the need to carefully design vegetarian meal options to minimize potential adverse effects.

Moreover, although MFD yields substantial reductions in GHG emissions associated with campus food consumption on treated days themselves, our findings identify {customer retention} as a critical barrier to the overall success of the intervention.
The identification of retention as a pivotal factor highlights two potential strategies for enhancing the environmental impact of MFD. The first involves integrating MFD with complementary interventions designed to improve customer participation in MFD on campus. The second approach suggests moving beyond rigid frameworks like MFD by intervening more deeply in the environmental setup of cafeterias. This could involve strategies such as reducing the prominence of high-emission ingredients while promoting a balanced selection of animal- and plant-based meal options.

\section*{Experimental Design}\label{sec:experiment_setup}

\xhdr{Meat-Free Day schedule} We implemented Meat-Free Day (MFD) in all 12 cafeterias of the EPFL campus for a period of 18 months (March 2022 to August 2023). On MFD, all meals served in the campus cafeterias—each offering between two and seven menu options—were exclusively plant-based.

Unlike traditional setups where MFD is scheduled on a specific day of the week (e.g., Meatless Monday~\cite{milford2019meat}), our approach consisted in assigning MFD to a random weekday for each week of the study, such that MFD was not associated with a specific weekday. The campus administration randomly selected one weekday per week throughout the study period, exluding weekends and public holidays. During the study period, a total of 77 MFDs were scheduled.

Importantly, the campus community—including students, employees, and visitors—was not notified in advance of MFDs. Participants could identify MFDs only via using the campus app that shows menus of the cafeterias (slightly before service time) or upon entering cafeterias, where digital screens displayed meal options that were entirely vegetarian. For our analysis, we focus on meals purchased during lunch, as the vast majority of the campus community consumes their primary meal at that time.

\xhdr{Estimation} To assess the impact of MFD, we employed two complementary experimental setups: \emph{On}-MFD Analysis and \emph{Beyond}-MFD Analysis.

In the \emph{on-MFD Analysis,} we used a matched-pairs design, comparing each MFD to a corresponding non-MFD from the preceding week. Both days were matched to fall on the same weekday (e.g., a meat-free Wednesday was compared to the previous week’s non-meat-free Wednesday; see Fig. \ref{fig:overview}(a)). This design accounted for day-specific patterns in cafeteria attendance. To further improve the comparability within matched pairs, we excluded week pairs where
(i) MFD fell on the same weekday in both weeks, (ii) one of the two weeks was the first or last week of a semester, or (iii) the two weeks fell in different periods of the academic calendar  (e.g., lecture weeks, exam periods, or breaks), to avoid comparing weeks with  different levels of on-campus presence.
This process yielded 67 valid MFD\slash non-MFD pairs, corresponding to 134 analyzed days and 402{,}536 food purchase transactions.

Using this matched sample, we estimated the variation ($\Delta$) in three key outcomes between MFDs and their matched non-MFDs. First, greenhouse gas emissions ($\Delta_{\text{GHG}}$), measured as the equivalent kilograms of CO$_2$ (kgCO$_2$e) associated with the entirety of meals sold daily. The CO$_2$ equivalent (kgCO$_2$e) is a standardized measure used to compare emissions from various greenhouse gases (e.g., methane) based on their global warming potential relative to the one of carbon dioxide. Second, the number of meals sold daily ($\Delta_{\text{sales}}$), obtained from EPFL’s transaction log system, which records real-time sales across all campus cafeterias; and (iii) the variation in caloric ($\Delta_{\text{cal}}$) and macronutrient ($\Delta_{\text{nutr}}$) intake per customer.

In the \emph{beyond-MFD Analysis,} we investigated whether MFD leads to sustained behavioral changes or unintended compensatory effects on the same three outcomes as above. We again employed a matched-pairs design, now comparing two non-MFDs from consecutive weeks (Fig. \ref{fig:overview}b): one non-MFD immediately following the MFD (week $t+1$), the other non-MFD occurring at least two days after the MFD (week $t$).  We obtained 56 valid MFD\slash non-MFD pairs for this setup.

We estimated the effect of MFD on the above outcomes using a generalized linear model (GLM) (see \textit{Materials and Methods} for model specification).

\begin{figure*}[t]
    \centering
    \includegraphics[width=1.0\textwidth]{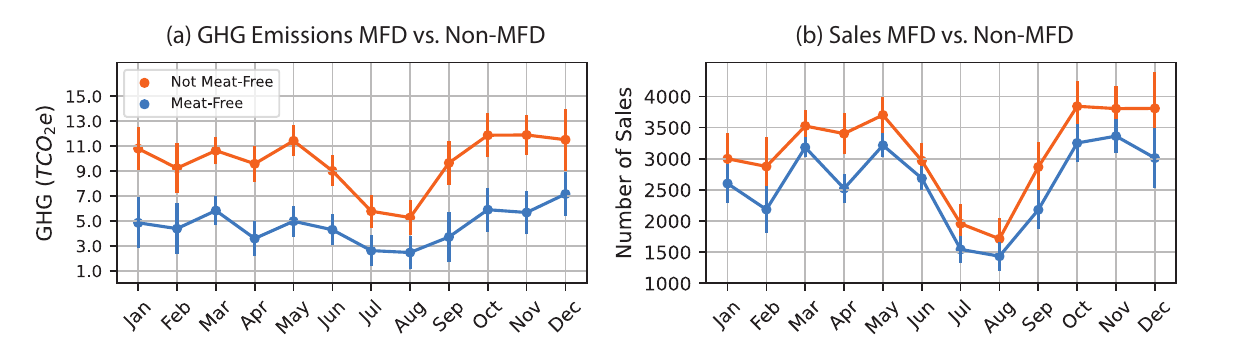}
\caption{\footnotesize \justifying \textbf{Impact of Meat-Free Days (MFD) on greenhouse gas (GHG) emissions and on-campus meal sales.}
\textbf{(a) GHG emissions:} Average monthly GHG emissions on MFDs (blue) compared to matched non-MFDs (orange). We observe a consistent and significant reduction in GHG emissions throughout the study period, with an overall decrease of 5,053 kgCO$_2$e/day ($p<0.001$), corresponding to a relative reduction of 52.9\%.
\textbf{(b) Sales:} Average monthly on-campus meal sales on MFDs (blue) compared to non-MFDs (orange). Meal sales significantly decreased by an average of 16.8\% ($p<0.001$) on MFDs, suggesting increased off-campus dining and potential offsetting of on-campus emission savings.}

    \label{fig:rq1_emissions}
    \vspace{-1mm}

\end{figure*}

\section*{Results of on-MFD Analysis}

\xhdr{MFD reduces immediate greenhouse gas emissions (RQ1)}\label{subsec:rq1_on_mfd} In order to analyze
the impact of MFD on greenhouse gas (GHG) emissions across the 12 campus cafeterias, we compared the total daily GHG emissions from meals sold on MFDs to those sold on matched non-MFDs (see Fig. \ref{fig:overview}(a)) using the regression model described in Equation \ref{eq:reg}. 
The model estimated a decrease in overall GHG emissions of $\Delta_{\text{GHG}} = -5\,053$ kgCO$_2$e per day ($p<0.001$, $t=-11.94$, 95\% CI $[-5\,889, -4\,216]$, see \emph{Table I in SI}),  from an average of $9\,534$ kgCO$_2$e per day ($p<0.001$, $t=31.87$, 95\% CI $[8\,942, 10\,100]$) on non-MFDs to an average $4{,}481$  per day ($p<0.001$, $t=14.98$, 95\% CI $[3\,889, 5\,072]$) on MFDs. This decrease corresponds to a relative decrease in terms of GHG emissions of 52.9\%. W

Fig. \ref{fig:rq1_emissions}(a) illustrates the average GHG emissions for both MFDs and non-MFDs by month of the year. The reduction in emissions remained consistent and statistically significant throughout all months of the study, suggesting that the effect of the intervention is robust to seasonal variations.

\xhdr{MFD increases off-campus dining (RQ2)}
It is conceivable that MFD can create a sense of restriction, causing some diners to purchase animal-based meals off-campus, thus  limiting the benefits of MFD. Since off-campus transaction data is not available to us, we instead approximate the impact of MFD on off-campus dining by analyzing whether MFD led to a measurable decrease in on-campus meal sales.

Estimating the average daily percentage change in food sales on MFDs compared to matched non-MFDs revealed a significant daily decrease in sales of $\Delta_\text{sales}=16.8\%$ on MFDs ($p<0.001$, $t=-11.92$, 95\% CI  $[-19.9\%, -14.2\%]$, see \emph{Table III in SI}) ). Daily sales declined from $3,092$ ($p<0.001$, $t=32.05$,  95\% CI $[2\,901$, $3\,283$])  on non-MFDs  to $2{,}579$ on MFDs ($p<0.001$, $t=-12.29$, 95\% CI $[2\,489, 2\,656]$). This  corresponds to an average daily reduction of 520 customers (under the assumption that each customer consumed one meal).

Fig. \ref{fig:rq1_emissions}(b) illustrates the  daily average on-campus sales for both MFDs and non-MFDs by  month of the year. The difference between meal purchases on MFDs vs.\ non-MFDs is statistically significant throughout the study period.

These results emphasize that, although MFD achieves significant immediate on-campus GHG reductions on the treated days themselves, overall effects may be reduced due to lacking customer retention. The following analysis addresses this concern.

\begin{figure}[t]
    \centering
    \includegraphics[width=0.9\columnwidth]{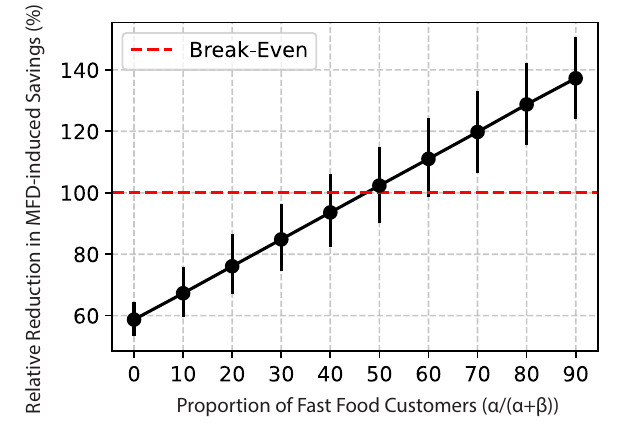}
     \caption{\footnotesize \justifying \textbf{Impact of off-campus dining on GHG emission savings.} Results from Monte Carlo simulations accounting for variability in the proportion of fast-food customers (x-axis). The x-axis represents the expected proportion of fast-food customers, computed as the mean $\alpha / (\alpha + \beta)$ of a Beta distribution  where $\alpha$ and $\beta$ represent the assumed number of fast-food and non-fast-food customers, respectively. The y-axis shows the percentage offset in GHG due to off-campus dining. In the best-case scenario, where no off-campus customers visit fast-food restaurants, the offset effect reaches 58.7\%. The average offset is not statistically different from the break-even point (red line) at 40\% (95\% CI [82.2\%, 106.1\%]). On average, the break-even point is reached at 50.2\% fast-food proportion, beyond which the offset exceeds the emissions savings from MFD.}

    \label{fig:tradeoff}
    \vspace{-1mm}

\end{figure}

\xhdr{Off-campus dining may negate environmental benefits (RQ1, RQ2)}
Off-campus diners can choose between two options: (i) a fast-food restaurant offering a discounted student menu and (ii) a mix of alternatives, including pre-packaged grocery store meals, a taco shop (partially closed during the study), and a higher-end sit-down restaurant (We provide a map of the campus showing the distance from campus to these options, see \emph{SI Section B}).

Using GHG emissions data from comparable on-campus meals, we estimate that a meal from the fast-food student menu generates an average of \(13.5\) kgCO\(_2\)e (SD = 3.5) per customer, while other off-campus options—assuming a typical meal structure of animal protein, dairy, carbohydrates, and a side—emit an average of \(5.7\) kgCO\(_2\)e (SD = 2.5) per meal.

Since the exact proportion of opt-out customers choosing fast food vs.\ other alternatives, as well as the GHG emissions from all off-campus meals, are unknown, we use Monte Carlo simulation to model the impact of off-campus dining under the variability of these parameters (see \textit{Materials and Methods} for details).

Under the most optimistic scenario—where no opt-out customers choose fast food—off-campus dining offsets 58.6\% (95\% CI [53.1\%, 64.4\%]) of the GHG reductions achieved by MFD (see Fig.~\ref{fig:tradeoff}). This scenario is, however, unlikely. The campus community is composed of roughly 70\% students, a group more likely to be drawn to the discounted student menu at the fast-food restaurant.
In Fig.~\ref{fig:tradeoff} we show that as the 
proportion of opting-out customers eating at the fast-food increases, the offset effect intensifies. At 40\% fast-food participation, off-campus emissions counteract 93.6\% (95\% CI $[82.2\%, 106.1\%]$) of MFD savings, reaching levels statistically indistinguishable from full offset (i.e., 100\%).

We estimate that off-campus emissions fully offset MFD-induced savings when 50.2\% of opt-out customers choose fast food. This ``break-even'' point corresponds to 270 customers, equivalent to 8.7\% of the campus population that  dines in cafeterias on a regular day. Beyond this threshold, off-campus dining not only negates the positive impact of MFD, but even leads to a net increase in emissions, surpassing levels observed on non-MFD days.

\xhdr{MFD maintains caloric intake but affects macronutrient mix (RQ3)}
The scope of MFD extends beyond sustainability, offering the prospect to also promote healthier dietary habits by improving the consumption of key nutrients. We therefore estimated the difference in the average caloric  ($\Delta_\text{cal}$) and macronutrient consumption ($\Delta_{\text{nutr}}$) per customer on MFDs vs.\ non-MFDs.

\begin{figure}[t]
    \centering
    \includegraphics[width=1.0\columnwidth]{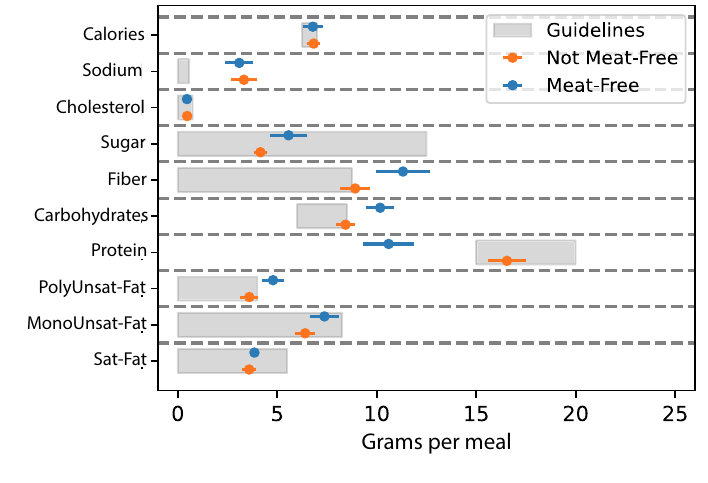}
\caption{\footnotesize \justifying \textbf{Impact of Meat-Free Day (MFD) on caloric and macronutrient intake.} 
Average intake of calories and macronutrients on MFDs and non-MFDs (x-axis) compared to recommended guidelines (gray boxes). Calories are measured in kcal and scaled by a factor of 100, whereas carbohydrates are scaled by a factor of 10 to align with the visualization of error bars for other macronutrients. This figure shows the diverse impact of MFD on nutrition: while it preserves caloric intake and it improves fiber and cholesterol consumption, there are risk due to lower protein and higher carbohydrate intake.}
    \label{fig:nutrition}
    \vspace{-1mm}

\end{figure}

We first assessed whether the daily caloric intake during non-MFDs complied with recommended dietary guidelines from governmental institutions~\cite{SwissDietaryRecommendations, FAOEuropeanDietaryGuidelines}. The analysis revealed an average caloric intake of 682.9 kcal per customer ($p < 0.001$, $t = 41.07$, 95\% CI $[650.0, 715.8]$, see \emph{Table V in SI}), aligning with the suggested range of 625--700 kcal per meal (see Fig.~\ref{fig:nutrition}, top row).

We then examined whether caloric intake differed between MFDs and non-MFDs. The results showed no statistically significant difference, with an estimated difference of $\Delta_{\text{cal}} = -4.1$ kcal ($p = 0.864$, $t = -0.171$, 95\% CI $[-50.53, 42.50]$, see \emph{Table V in SI}). These findings indicate that MFDs do not alter the caloric intake per customer.

Macronutrient consumption on non-MFDs  aligns with recommended dietary guidelines, with the sole exception of sodium. During non-MFDs, the average sodium intake is 3.32 grams per meal ($p < 0.001$, $t = 9.86$, 95\% CI $[2.6, 3.9]$, see \emph{Table S5 in SI}), exceeding the recommended limit of 1.6 grams per meal. MFD reduction of sodium remain not statistically significant different from the non-MFD sodium consumption ($\Delta_{\text{sod}} = -0.23$ grams ($p=0.626$, $t = -0.48$, 95\% CI $[-1.18, 0.71]$) and therefore  remains above guideline levels (see Fig.~\ref{fig:nutrition}, second row).

MFDs improved the nutritional profile of meals consumed by increasing fiber intake and reducing cholesterol intake. Specifically, fiber consumption increased by $\Delta_{\text{fiber}} = 2.40$ grams per meal ($p < 0.001$, $t = 4.44$, 95\% CI $[1.33, 3.47]$, see \emph{Table V in SI}), representing a 26.9\% increase from the average fiber intake of 8.91 grams on non-MFDs. Similarly, cholesterol consumption decreased by $\Delta_{\text{chol}} = -0.23$ grams per meal ($p < 0.005$, $t = -0.313$, 95\% CI  $[-0.36, -0.09]$, , see \emph{Table V in SI}), corresponding to a 4.50\% reduction. Furthermore, monounsaturated fat intake increased by $\Delta_{\text{M-fat}} = 0.97$ grams per meal ($p = 0.009
$, $t = 2.64$, 95\% CI $[0.26, 1.70]$) corresponding to a relative increase of 15.2\%.
These variations are desirable, as they are associated with reduced risk of cardiovascular disease~\cite{anderson2009health, anderson1994health} and  improvements for digestive health~\cite{silverman2016association, gillingham2011dietary}.

On the flip side, MFDs resulted in a significant reduction in protein intake (\( \Delta_{\text{protein}} = -5.94 \) grams, \( p < 0.001 \), \( t = -8.76 \), 95\% CI $[-7.29, -4.60]$, see \emph{Table V in SI}), and in a increase in carbohydrates consumption \( \Delta_{\text{carbs}} = 17.38 \) grams, \( p < 0.001 \), \( t = 4.98 \), 95\% CI $[10.47, 24.30]$). These variations corresponded to a 27.6\% decrease for proteins and a 20.6\% increase for carbs pushing the consumption of these nutrients beyond the quantities reccomended by guidelines. Also, sugar intake increased by \( \Delta_{\text{sugar}} = 1.42 \) grams (\( p < 0.001 \), \( t = 5.99 \), 95\% CI  $[0.95, 1.88]$), representing a 34.2\% increase, but its consumption stays within recommended guidelines. These changes indicate possible nutritional shortfall, given the critical role of protein in maintaining muscle mass \cite{phillips2012nutrient}, the risk of fluctuations in blood sugar \cite{mantantzis2019sugar}, and potential weight gain \cite{van2007carbohydrate}.

\xhdr{MFD does not affect unhealthy-item consumption (RQ3)} The perception that vegetarian meals lack sufficient nutrition or energy is well-documented \cite{appleton2016barriers, jahn2021plant, kemper2020motivations}. This perception may lead individuals to compensate by purchasing additional, potentially unhealthy items on MFDs. To investigate this, we analyzed the percentage daily changes in sales across four categories: (i) vending machine items, (ii) energy and sugary drinks, (iii) pastries and snacks, and (iv) desserts.

\begin{figure}[t]
    \centering
    \includegraphics[width=1.0\columnwidth]{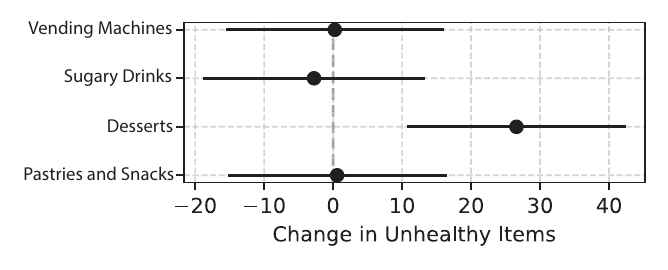}
\caption{\footnotesize \justifying \textbf{Changes in sales of unhealthy food items on Meat-Free Days (MFDs).} Percentage daily changes in sales for vending machine items, energy and sugary drinks, pastries and snacks, and desserts on MFDs relative to non-MFDs. No significant differences were observed for vending machine items, energy and sugary drinks, or pastries and snacks. However, dessert sales increased by 15.9\% ($p<0.001$), corresponding to an additional 27 desserts sold compared to a baseline of 167 desserts on non-MFDs. This increase suggests a potential compensatory behavior driven by the perception that vegetarian meals are less nutritionally adequate.}
    \label{fig:nutrition_unhealthy}
    \vspace{-1mm}

\end{figure}

Our analysis revealed no statistically significant changes in sales for vending machine items, energy and sugary drinks, or pastries and snacks (see Fig.~\ref{fig:nutrition_unhealthy}). Dessert sales, however, increased  on MFDs, with an estimated percentage increase per customer of \( \Delta_{\text{sales}} = 15.9\% \) (\( p < 0.001 \), \( t = 3.35 \), 95\% CI $[10.9, 27.2]$). This corresponds to an additional 27 desserts sold compared to a baseline of 167 desserts on non-MFDs (\( p < 0.001 \), \( t = 22.16 \), 95\% CI $[151.5, 180.9]$).

Since desserts are  purchased at the same time as meals, customers may already be aware that they will be eating a vegeterian meal. This awareness could trigger pre-existing beliefs that vegetarian meals are less nutritionally adequate, prompting them to complement their meal with a dessert. In contrast, the other three food categories are usually bought after dining, when the main meal choice has already been made, and are therefore less influenced by the perceived nutritional content of the meal.
This may explain why desserts are affected by the intervention, while other categories are not.

\section*{Results of beyond-MFD Analysis}

\xhdr{MFD triggers rebound in GHG emissions (RQ1)} The absence of animal-based options on MFDs may trigger, among some diners, a rebound effect that increases meat consumption on subsequent days. To assess this effect, we used our second Setup \emph{Beyond MFD} defined in \textit{Experimental Design} Section (see Fig. \ref{fig:overview}).

Our analysis revealed a statistically significant increase in GHG emissions of 414.2~kgCO$_2$e ($p < 0.001$, $t = 8.68$, 95\% CI  $[366.5, 461.9]$) on the non-MFD day immediately following an MFD  compared to a non-MFD day at least two days after the MFD. Accounting for this rebound effect, the net daily GHG savings attributed to MFD are reduced to $\Delta_{\text{GHG}} = -4\,634~\text{kgCO}_2\text{e}$, down from the initial 5,053~kgCO$_2$e estimated in the on-MFD Analysis (\emph{on-MFD Analysis}, RQ1). We report all results in \textit{Section A Table I and II of the SI}

\xhdr{MFD triggers rebound in animal-based meal sales (RQ2)} Using the beyond-MFD Analysis setup, we examined changes in animal-based meal consumption, defining these meals as those containing white or red meat, fish, seafood, or dairy as primary protein sources. Our analysis identified a modest but significant increase of $\Delta_{\text{sales}}=3.5\%$ ($p=0.008$, $t=2.70$, 95\% CI $[0.7\%, 5.8\%]$). Consumption rose from 56.8\% on non-MFDs at least two days after an MFD ($p<0.001$, $t=63.26$, 95\% CI $[55.2\%, 58.8\%]$) to 60.5\% ($p<0.001$, $t=66.55$, 95\% CI $[58.7\%, 62.3\%]$) on non-MFDs immediately following an MFD.

\xhdr{MFD does not trigger nutritional rebound (RQ3)} To assess whether MFDs have a spillover effect on customers’ nutritional intake on subsequent non-MFDs, we analyzed variations in caloric and macronutrient consumption ($\Delta_{\text{cal}}$, $\Delta_{\text{nutr}}$) following MFDs.
Our analysis did not identify any statistically significant changes in overall caloric intake or macronutrient composition between non-MFDs occurring one day after an MFD and those occurring more than one day after an MFD. The estimated variation in caloric intake was $\Delta_{\text{cal}} = 2.4$ kcal ($p = 0.567$, 95\% CI $[-78.2, 82.3]$), with similar null results observed across all macronutrient categories. Likewise, we found no statistically significant changes in the consumption of vending machine items, sugary drinks, desserts, or pastries.


\section*{Robustness Checks}

To ensure the validity of our findings, we conducted two robustness checks assessing potential biases in our experimental design

\xhdr{Placebo test} We first conducted a placebo test by comparing two non-MFD days occurring on the same weekday in consecutive weeks (similarly to Fig. \ref{fig:overview}(b) but without considering the distance from the MFD). 
Using the same constraints as the main analysis (see Estimation in the Experimental Design Section, we collected 260 matched non-MFD pairs. The results confirm the absence of systematic differences, with no significant variation in GHG emissions (92.32 KgCO$_2$e, $t = 0.12$, $p = 0.901$) or meal sales ($-57$ purchases, $t = -0.27$, $p = 0.788$). Macronutrient intake also showed no statistically significant differences across all tested categories (see Supplementary Information)
This test  confirms that there are no differences between same weekdays of consecutive weeks in terms of greenhouse gas (GHG) emissions, sales, and nutritional intake. Therefore, the observed difference of our study are determined by MFD.
.

\xhdr{Comparison with subsequent week} In our primary analysis, we matched each MFD with a non-MFD from the previous week. To test the robustness of our results, we repeated the analysis using a non-MFD from the following week instead. Our results remain identical compared to those shown in our main analysis, and significantly so. More specifically, for RQ1 and RQ2 in the on-MFD Analysis, we found a reduction of GHG emissions of $5\,338$ kgCO$_2$e ($p<0.001$, $t=-14.98$, 95\% CI $[-6\,167, -4\,508]$). Also, we observed a reduction of 531 transactions ($p=0.04$, $t= -2.56$,  95\% CI $[-937, -125]$).
For RQ1 and RQ2 in the {beyond}-MFD analysis, we find an increase in GHG emissions of 431 kgCO$_2$e ($p=0.012$, $t=11.62$, 95\% CI $[358, 504]$) and an increase in animal-based meals of 3.2\% ($p=0.006$, $t=2.50$, 95\% CI $[0.1\%, 5\%]$). Results on nutritional values remain statistically significant equivalent to those of our main analysis, we report this in the Supplementary materials given the high number of  macronutrients that we analyzed in our analysis.

\section*{Discussion}
Our study demonstrates the potential of Meat-Free Day (MFD) in advancing more sustainable and healthier food systems. By reducing immediate on-campus greenhouse gas (GHG) emissions by over 50\%, MFD emerges as an easy-to-implement strategy to mitigate the environmental impact of campus dining. Beyond curbing GHG emissions, MFD also improves dietary quality on key metrics—boosting fiber intake and lowering cholesterol—all without affecting caloric intake.

Our findings, nonetheless, highlight the vulnerability of such interventions to behavioral spillover. If 8.7\% of regular cafeteria customers choose to eat high-emission meals off-campus—rather than participating in MFD—it could entirely nullify the GHG reductions achieved through MFD. We also observed a 3.52\% increase in animal-based meal consumption 6on non-MFD days following an MFD, compared to other non-MFD days. This pattern suggests that strict interventions like MFD, which force customers to choose plant-based meals, may inadvertently push consumers toward less sustainable choices when restrictions are lifted. This dynamic raises concerns about the effectiveness of MFD in achieving sustained reductions in institutional emissions and in catalyzing lasting dietary shifts.

\xhdr{Policy implications} Our analysis identifies {customer retention} as the central challenge limiting the efficacy of MFD and hints at two alternative pathways for addressing this issue.

The first pathway emphasizes synergistic approaches to incentivize customer retention on campus during MFD. Loyalty programs, such as fidelity cards, present a promising avenue for boosting customer retention. These cards grant discounts on subsequent meals after participating on MFDs, they require minimal logistical changes, and have proven effective in fostering customer retention in similar contexts \cite{berman2006developing, gold2024exploring}. 

The second pathway proposes flexible interventions that reduce the rigidity of MFD and promote gradual dietary changes. Strictly removing all animal-based meals provokes resistance and unintended behaviors, such as the increase in off-campus dining. Instead, targeted strategies—like reducing the availability of high-emission foods such as red meat and substituting them with more sustainable options (white meat, fish)—can achieve GHG reductions without alienating diners. These less restrictive measures align with behavioral research, which highlights the effectiveness of subtle modifications to environments in steering consumer choices \cite{garnett2019impact, handziuk2023carbon}.

\xhdr{Future work and limitations} Future research should focus on testing MFD in combination with other interventions to identify the most effective strategies for improving customer retention and unlocking the full environmental benefits of these initiatives. Synergistic strategies, such as loyalty programs or targeted reductions in high-emission foods, should be examined to understand how they interact with MFD in fostering on-campus engagement and reducing behavioral spillover effects.

Further studies should also address a critical limitation of our analysis: the inability to monitor food behaviors off-campus. Whereas we have robust data about on-campus dining, we lack information on whether diners compensate by consuming high-emission meals elsewhere. This limitation could be addressed through complementary methods such as interviews, surveys, or data donation studies, which would provide a more comprehensive understanding of rebound effects and allow for more accurate evaluations of MFD efficacy.

Additionally, exploring the implementation of MFD in diverse settings, such as hospitals, workplaces, or community dining spaces, could provide valuable insights into how different environments influence their effectiveness. These contexts may exhibit unique dining behaviors and logistical challenges, which could inform the design of tailored interventions to maximize retention and impact.

\section*{Materials and Methods}\label{sec:MatandMeth}

\xhdr{Data} We leverage an anonymized dataset of food purchases made on the EPFL university campus. The data span from February 1, 2022, to September 1, 2023, and contain approximately 3.3 million transactions. Each transaction is linked to an anonymized ID and includes metadata such as the timestamp, location, cash register, purchased items, and price. The dataset covers all food outlets permanently located on campus, including restaurants, cafes, and vending machines. The locations are illustrated in the Supplementary Material, Section S1.7. All transactions are annotate with a textual description representing the category associated with the transaction  (e.g., dessert, snack, student menu). 

For transactions occurring during lunch hours (11:00–14:00), we integrate additional information, including a detailed textual description of the meal, a breakdown of its ingredients, the greenhouse gas (GHG) emissions per ingredient, and portion sizes (i.e., quantity of each ingredient). The dataset comprises over 7,700 unique ingredients. Examples of meal options available on Meat Free Days (MFDs) and non-MFDs are provided in the Supplementary Material.

Additionally, each transaction is annotated with demographic attributes, including gender, age, and campus status (student, staff, visitor). Campus meal prices range from CHF 5.5 to CHF 16.5, with discounts available for students and staff. Our analysis of Meat Free Days (MFDs) focuses on 363,630 meal purchases. For unhealthy items such as snacks and drinks consumed outside main meals, we extend the timeframe and examine 402,536 transactions. A detailed breakdown is provided in the Supplementary Information (SI).

\xhdr{Statistical model}  

To analyze the effect of Meat-Free Days (MFD) on various outcomes, we formalize our model as follows:

\begin{equation}
y_i = \alpha + \beta \cdot A_i + \beta_1 \cdot P_i + \epsilon_i
\label{eq:reg}
\end{equation}

where \( A_i \in \{0,1\} \) indicates whether day \( i \) is a Meat-Free Day, and \( P_i \) is a categorical variable representing the period of the academic calendar to which the day belongs (lecture, break, or exam). The inclusion of \( P_i \) accounts for potential seasonal effects or behavioral shifts in food consumption across different academic periods. The error term \( \epsilon_i \) captures unobserved factors influencing the outcome.

The outcome variable \( y_i \) varies across RQs.  \( y_i \) represents (i) the total GHG emissions associated with all meals (RQ1) sold on a given day,  (ii) the total number of meals sold (RQ2), and the pverall daily caloric and macronutrients intake (RQ3).

\xhdr{Monte Carlo simulation of off-campus GHG emissions} To estimate the greenhouse gas (GHG) emissions associated with off-campus dining on meat-free days (MFDs), we accounted for two key uncertainties: the proportion of opting-out diners who choose high-emission fast-food restaurants versus lower-emission alternatives, and the emissions associated with these meals, given the absence of direct sales transaction data.

To address this, we implemented a Monte Carlo simulation to explore a range of plausible scenarios and assess whether the emissions savings from MFDs are sustained or offset under different behavioral assumptions. The simulation is parameterized by three components: the number of diners opting out of on-campus meals, the proportion selecting fast-food restaurants, and the emissions per meal in each dining category. 

The number of opting-out diners is modeled as a Poisson-distributed variable with \( \lambda = 520 \), based on prior estimates. The fraction of diners choosing fast food is drawn from a Beta distribution, \( F(x; \alpha, \beta) \), where \( \alpha \) and \( \beta \) determine the shape of the distribution and control the likelihood of different fast-food uptake proportions. Larger values of \( \alpha \) relative to \( \beta \) increase the probability of diners selecting fast food, while larger values of \( \beta \) bias the distribution toward alternative dining choices. 

To account for variability in emissions, meal-related GHG emissions are modeled as normally distributed, where fast-food meals follow \( \mathcal{N}(14.2, 3.5^2) \) kgCO\(_2\)e per meal, and other alternatives follow \( \mathcal{N}(5.7, 2.5^2) \) kgCO\(_2\)e per meal.

For each simulation iteration, the number of opting-out diners is drawn from a Poisson distribution. Each diner is assigned to fast food or an alternative dining option using the Beta-distributed fraction. Emissions are then sampled from the respective normal distributions and summed to estimate total off-campus emissions. Net emissions are computed by subtracting the estimated MFD savings of 5,053.02 kgCO\(_2\)e. This process is repeated 10,000 times for each Beta parameter configuration to generate a distribution of net emissions outcomes.
From each set of 10,000 simulations, we compute the mean net emissions and the associated 95\% confidence intervals (2.5th and 97.5th percentiles). We also track whether net emissions exceed zero, identifying the proportion of fast-food diners at which MFD savings are fully offset.

\clearpage
\onecolumngrid

\bibliographystyle{plain}
\bibliography{refs}

\begin{thebibliography}{10}

\bibitem{anderson2009health}
James~W Anderson, Pat Baird, Richard~H Davis~Jr, Stefanie Ferreri, Mary Knudtson, Ashraf Koraym, Valerie Waters, and Christine~L Williams.
\newblock Health benefits of dietary fiber.
\newblock {\em Nutrition reviews}, 67(4):188--205, 2009.

\bibitem{anderson1994health}
James~W Anderson, Belinda~M Smith, and Nancy~J Gustafson.
\newblock Health benefits and practical aspects of high-fiber diets.
\newblock {\em The American journal of clinical nutrition}, 59(5):1242S--1247S, 1994.

\bibitem{appleton2016barriers}
KM~Appleton.
\newblock Barriers to and facilitators of the consumption of animal-based protein-rich foods in older adults.
\newblock {\em Nutrients}, 8(4):187, 2016.

\bibitem{berman2006developing}
Barry Berman.
\newblock Developing an effective customer loyalty program.
\newblock {\em California management review}, 49(1):123--148, 2006.

\bibitem{blondin2022meatless}
Stacy~A Blondin, Sean~B Cash, Timothy~S Griffin, Jeanne~P Goldberg, and Christina~D Economos.
\newblock Meatless monday national school meal program evaluation: impact on nutrition, cost, and sustainability.
\newblock {\em Journal of Hunger \& Environmental Nutrition}, 17(1):1--13, 2022.

\bibitem{brehm1966theory}
Jack~W Brehm.
\newblock {\em A theory of psychological reactance.}
\newblock Academic press, 1966.

\bibitem{nyt_meatout_2021}
Alison Brennan.
\newblock Meatout day sparks debate between colorado and nebraska.
\newblock \url{https://www.nytimes.com/2021/03/16/us/meatout-day-colorado-nebraska.html}, 2021.
\newblock Accessed: 2024-11-25.

\bibitem{veggie_day_ghent}
{City of Ghent}.
\newblock Ghent’s veggie day: Pioneering meat-free days in europe, 2009.
\newblock Accessed: 2024-08-09.

\bibitem{craig2021safe}
Winston~J Craig, Ann~Reed Mangels, Uju{\'e} Fres{\'a}n, Kate Marsh, Fayth~L Miles, Angela~V Saunders, Ella~H Haddad, Celine~E Heskey, Patricia Johnston, Enette Larson-Meyer, et~al.
\newblock The safe and effective use of plant-based diets with guidelines for health professionals.
\newblock {\em Nutrients}, 13(11):4144, 2021.

\bibitem{earthday2019}
Earth Day.
\newblock Un report: Plant-based diets provide “major opportunities” to address climate crisis.
\newblock {\em Earth Day}, 2019.

\bibitem{deboer2014attitudes}
Joop De~Boer, Hanna Schosler, and Harry Aiking.
\newblock Attitudes towards meat-free days: An exploratory study among dutch consumers.
\newblock {\em Appetite}, 76:113--123, 2014.

\bibitem{de2012nutritional}
Willem De~Keyzer, Sven Van~Caneghem, Anne-Louise~M Heath, Barbara Vanaelst, Mia Verschraegen, Stefaan De~Henauw, and Inge Huybrechts.
\newblock Nutritional quality and acceptability of a weekly vegetarian lunch in primary-school canteens in ghent, belgium:‘thursday veggie day’.
\newblock {\em Public health nutrition}, 15(12):2326--2330, 2012.

\bibitem{erfanian2024cultivating}
Sahar Erfanian, Shengze Qin, Liaqat~Ali Waseem, and Muneer~Ahmed Dayo.
\newblock Cultivating a greener plate: understanding consumer choices in the plant-based meat revolution for sustainable diets.
\newblock {\em Frontiers in Sustainable Food Systems}, 7:1315448, 2024.

\bibitem{estell2021plant}
Madeline Estell, Jaimee Hughes, and Sara Grafenauer.
\newblock Plant protein and plant-based meat alternatives: Consumer and nutrition professional attitudes and perceptions.
\newblock {\em Sustainability}, 13(3):1478, 2021.

\bibitem{SwissDietaryRecommendations}
{Federal Food Safety and Veterinary Office}.
\newblock {Swiss Dietary Recommendations}, n.d.
\newblock Accessed: 2025-01-17.

\bibitem{feher2020comprehensive}
Andr{\'a}s Feh{\'e}r, Micha{\l} Gazdecki, Mikl{\'o}s V{\'e}ha, M{\'a}rk Szak{\'a}ly, and Zolt{\'a}n Szak{\'a}ly.
\newblock A comprehensive review of the benefits of and the barriers to the switch to a plant-based diet.
\newblock {\em Sustainability}, 12(10), 2020.

\bibitem{FAOEuropeanDietaryGuidelines}
{Food and Agriculture Organization of the United Nations (FAO)}.
\newblock {European Dietary Guidelines}, n.d.
\newblock Accessed: 2025-01-17.

\bibitem{oecd2021meat}
Organisation for Economic Co-operation, Development (OECD), Food, and Agriculture~Organization (FAO).
\newblock {\em OECD-FAO Agricultural Outlook 2021-2030}, chapter Chapter Six: Meat.
\newblock OECD Publishing, 2021.

\bibitem{frontiers2021}
Frontiers.
\newblock Environmental impact of animal-based food production and the feasibility of a shift toward sustainable plant-based diets in the united states.
\newblock {\em Frontiers in Nutrition}, 2021.

\bibitem{garnett2019impact}
Emma~E Garnett, Andrew Balmford, Chris Sandbrook, Mark~A Pilling, and Theresa~M Marteau.
\newblock Impact of increasing vegetarian availability on meal selection and sales in cafeterias.
\newblock {\em Proceedings of the National Academy of Sciences}, 116(42):20923--20929, 2019.

\bibitem{giacalone2022understanding}
Davide Giacalone, Mathias~P Clausen, and Sara~R Jaeger.
\newblock Understanding barriers to consumption of plant-based foods and beverages: Insights from sensory and consumer science.
\newblock {\em Current Opinion in Food Science}, 48:100919, 2022.

\bibitem{gillingham2011dietary}
Leah~G Gillingham, Sydney Harris-Janz, and Peter~JH Jones.
\newblock Dietary monounsaturated fatty acids are protective against metabolic syndrome and cardiovascular disease risk factors.
\newblock {\em Lipids}, 46:209--228, 2011.

\bibitem{gligoric2024measuring}
Kristina Gligori{\'c}, Robin Zbinden, Arnaud Chiolero, Emre K{\i}c{\i}man, Ryen~W White, Eric Horvitz, and Robert West.
\newblock Measuring and shaping the nutritional environment via food sales logs: case studies of campus-wide food choice and a call to action.
\newblock {\em Frontiers in Nutrition}, 11:1231070, 2024.

\bibitem{gold2024exploring}
Natalie Gold, Pieter Cornel, Shi Zhuo, Katie Thornton, Rupert Riddle, and Robert McPhedran.
\newblock Exploring the impact of giving free food samples and loyalty cards on sustainable food choices: A stepped wedge trial in workplace food outlets.
\newblock {\em International Journal of Market Research}, 66(5):650--673, 2024.

\bibitem{handziuk2023carbon}
Yurii Handziuk and Stefano Lovo.
\newblock Carbon information, pricing, and bans. evidence from a field experiment.
\newblock {\em HEC Paris Research Paper No. FIN-2023-1493}, 2023.

\bibitem{congress_meatless_monday}
John Harris.
\newblock Meatless monday debate reaches u.s. congress, 2013.

\bibitem{hemler2019plant}
Elena~C Hemler and Frank~B Hu.
\newblock Plant-based diets for personal, population, and planetary health.
\newblock {\em Advances in Nutrition}, 10:S275--S283, 2019.

\bibitem{henry2019role}
Roslyn~C Henry, Peter Alexander, Sam Rabin, Peter Anthoni, Mark~DA Rounsevell, and Almut Arneth.
\newblock The role of global dietary transitions for safeguarding biodiversity.
\newblock {\em Global Environmental Change}, 58:101956, 2019.

\bibitem{jahn2021plant}
Steffen Jahn, Pia Furchheim, and Anna-Maria Str{\"a}ssner.
\newblock Plant-based meat alternatives: Motivational adoption barriers and solutions.
\newblock {\em Sustainability}, 13(23):13271, 2021.

\bibitem{kahleova2018vegetarian}
Hana Kahleova, Susan Levin, and Neal~D Barnard.
\newblock Vegetarian dietary patterns and cardiovascular disease.
\newblock {\em Progress in cardiovascular diseases}, 61(1):54--61, 2018.

\bibitem{kemper2020motivations}
Joya~A Kemper.
\newblock Motivations, barriers, and strategies for meat reduction at different family lifecycle stages.
\newblock {\em Appetite}, 150:104644, 2020.

\bibitem{klink2022socioeconomic}
Urte Klink, Jutta Mata, Roland Frank, and Benjamin Sch{\"u}z.
\newblock Socioeconomic differences in animal food consumption: Education rather than income makes a difference.
\newblock {\em Frontiers in Nutrition}, 9:993379, 2022.

\bibitem{lambrecht2023limiting}
Nathalie~J Lambrecht, Lesli Hoey, Alex Bryan, Martin Heller, and Andrew~D Jones.
\newblock Limiting red meat availability in a university food service setting reduces food-related greenhouse gas emissions by one-third.
\newblock {\em Climatic Change}, 176(6):67, 2023.

\bibitem{latona2024ai}
Giuseppe~Russo Latona, Manoel~Horta Ribeiro, Tim~R Davidson, Veniamin Veselovsky, and Robert West.
\newblock The ai review lottery: Widespread ai-assisted peer reviews boost paper scores and acceptance rates.
\newblock {\em arXiv preprint arXiv:2405.02150}, 2024.

\bibitem{lombardini2013forced}
Chiara Lombardini and Leena Lankoski.
\newblock Forced choice restriction in promoting sustainable food consumption: Intended and unintended effects of the mandatory vegetarian day in helsinki schools.
\newblock {\em Journal of consumer policy}, 36:159--178, 2013.

\bibitem{mantantzis2019sugar}
Konstantinos Mantantzis, Friederike Schlaghecken, Sandra~I S{\"u}nram-Lea, and Elizabeth~A Maylor.
\newblock Sugar rush or sugar crash? a meta-analysis of carbohydrate effects on mood.
\newblock {\em Neuroscience \& Biobehavioral Reviews}, 101:45--67, 2019.

\bibitem{politico_mfd_political}
Jane Martin.
\newblock Meatless monday becomes political battleground, 2013.
\newblock Accessed: 2024-08-09.

\bibitem{meatless_monday_origin}
{Meatless Monday}.
\newblock Meatless monday: A simple way to reduce your carbon footprint, 2024.
\newblock Accessed: 2024-08-09.

\bibitem{merk2024no}
Christine Merk, Leonie~P Meissner, Amelie Griesoph, Stefan Hoffmann, Ulrich Schmidt, and Katrin Rehdanz.
\newblock No need for meat as most customers do not leave canteens on veggie days.
\newblock {\em npj Climate Action}, 3(1):79, 2024.

\bibitem{dw_veggie_day_debate}
Franka Meyer.
\newblock Germany's veggie day proposal sparks political controversy, 2013.
\newblock Accessed: 2024-08-09.

\bibitem{micha2010red}
Renata Micha, Sarah~K Wallace, and Dariush Mozaffarian.
\newblock Red and processed meat consumption and risk of incident coronary heart disease, stroke, and diabetes mellitus: a systematic review and meta-analysis.
\newblock {\em Circulation}, 121(21):2271--2283, 2010.

\bibitem{milford2019meat}
Anna~Birgitte Milford and Charlotte Kildal.
\newblock Meat reduction by force: The case of “meatless monday” in the norwegian armed forces.
\newblock {\em Sustainability}, 11(10):2741, 2019.

\bibitem{un2022nutrition}
United Nations.
\newblock Un nutrition strategy 2022-2030, 2022.

\bibitem{un2023sdg}
United Nations.
\newblock The sustainable development goals report 2023, 2023.

\bibitem{academy2016}
Academy of~Nutrition and Dietetics.
\newblock Position of the academy of nutrition and dietetics: Vegetarian diets.
\newblock {\em Academy of Nutrition and Dietetics}, 2016.

\bibitem{parlasca2022meat}
Martin~C Parlasca and Matin Qaim.
\newblock Meat consumption and sustainability.
\newblock {\em Annual Review of Resource Economics}, 14(1):17--41, 2022.

\bibitem{phillips2012nutrient}
Stuart~M Phillips.
\newblock Nutrient-rich meat proteins in offsetting age-related muscle loss.
\newblock {\em Meat science}, 92(3):174--178, 2012.

\bibitem{rothgerber2013real}
Hank Rothgerber.
\newblock Real men don’t eat (vegetable) quiche: Masculinity and the justification of meat consumption.
\newblock {\em Psychology of Men \& Masculinity}, 14(4):363, 2013.

\bibitem{russo2024stranger}
Giuseppe Russo, Manoel~Horta Ribeiro, and Robert West.
\newblock Stranger danger! cross-community interactions with fringe users increase the growth of fringe communities on reddit.
\newblock In {\em Proceedings of the International AAAI Conference on Web and Social Media}, volume~18, pages 1342--1353, 2024.

\bibitem{russo2023spillover}
Giuseppe Russo, Luca Verginer, Manoel~Horta Ribeiro, and Giona Casiraghi.
\newblock Spillover of antisocial behavior from fringe platforms: The unintended consequences of community banning.
\newblock In {\em Proceedings of the international AAAI conference on web and social media}, volume~17, pages 742--753, 2023.

\bibitem{forbes_meatless_hospitals}
Joseph Sax.
\newblock Why hospitals and schools are going meatless on monday, 2019.
\newblock Accessed: 2024-08-09.

\bibitem{german_parliament_meatless_day}
Laura Schmidt.
\newblock German parliament debates "veggie day" proposal amid controversy, 2013.

\bibitem{schuz2021equity}
Benjamin Sch{\"u}z, Hannah Meyerhof, Lisa~Karla Hilz, and Jutta Mata.
\newblock Equity effects of dietary nudging field experiments: systematic review.
\newblock {\em Frontiers in Public Health}, 9:668998, 2021.

\bibitem{sciencedirect2019}
ScienceDirect.
\newblock Food in the anthropocene: the eat–lancet commission on healthy diets from sustainable food systems.
\newblock {\em ScienceDirect}, 2019.

\bibitem{silverman2016association}
Michael~G Silverman, Brian~A Ference, Kyungah Im, Stephen~D Wiviott, Robert~P Giugliano, Scott~M Grundy, Eugene Braunwald, and Marc~S Sabatine.
\newblock Association between lowering ldl-c and cardiovascular risk reduction among different therapeutic interventions: a systematic review and meta-analysis.
\newblock {\em Jama}, 316(12):1289--1297, 2016.

\bibitem{meat_free_movement}
John Smith.
\newblock The global meat-free movement: Over 40 countries on board, 2023.
\newblock Accessed: 2024-08-09.

\bibitem{springmann2018options}
Marco Springmann, Michael Clark, Daniel Mason-D’Croz, Keith Wiebe, Benjamin~Leon Bodirsky, Luis Lassaletta, Wim De~Vries, Sonja~J Vermeulen, Mario Herrero, Kimberly~M Carlson, et~al.
\newblock Options for keeping the food system within environmental limits.
\newblock {\em Nature}, 562(7728):519--525, 2018.

\bibitem{nytimes_meatless_monday}
Stephanie Strom.
\newblock Meatless monday catches on in schools and hospitals, 2012.
\newblock Accessed: 2024-08-09.

\bibitem{tilman2014}
David Tilman and Michael Clark.
\newblock Dietary guidelines and planetary health: Aligning recommendations for health and sustainability.
\newblock {\em PNAS}, 2014.

\bibitem{harvard_meatless}
Harvard University.
\newblock Harvard adds “meatless mondays” to campus dining, 2018.
\newblock Accessed: 2024-08-09.

\bibitem{van2007carbohydrate}
RM~Van~Dam and JC~Seidell.
\newblock Carbohydrate intake and obesity.
\newblock {\em European journal of clinical nutrition}, 61(1):S75--S99, 2007.

\bibitem{vandenbroele2020nudging}
Jolien Vandenbroele, Iris Vermeir, Maggie Geuens, Hendrik Slabbinck, and Anneleen Van~Kerckhove.
\newblock Nudging to get our food choices on a sustainable track.
\newblock {\em Proceedings of the Nutrition Society}, 79(1):133--146, 2020.

\bibitem{varela2022meat}
Paula Varela, Ga{\"e}lle Arvisenet, Antje Gonera, Kristine~S Myhrer, Viridiana Fifi, and Dominique Valentin.
\newblock Meat replacer? no thanks! the clash between naturalness and processing: An explorative study of the perception of plant-based foods.
\newblock {\em Appetite}, 169:105793, 2022.

\bibitem{time_meatless_monday}
Bryan Walsh.
\newblock Los angeles declares 'meatless mondays', 2012.
\newblock Accessed: 2024-08-09.

\bibitem{wang2023vegetarian}
Tian Wang, Andrius Masedunskas, Walter~C Willett, and Luigi Fontana.
\newblock Vegetarian and vegan diets: benefits and drawbacks.
\newblock {\em European heart journal}, 44(36):3423--3439, 2023.

\bibitem{willett2019food}
Walter Willett, Johan Rockstr{\"o}m, Brent Loken, Marco Springmann, Tim Lang, Sonja Vermeulen, Tara Garnett, David Tilman, Fabrice DeClerck, Amanda Wood, et~al.
\newblock Food in the anthropocene: the eat--lancet commission on healthy diets from sustainable food systems.
\newblock {\em The lancet}, 393(10170):447--492, 2019.

\bibitem{xu2021global}
Xiaoming Xu, Prateek Sharma, Shijie Shu, Tzu-Shun Lin, Philippe Ciais, Francesco~N Tubiello, Pete Smith, Nelson Campbell, and Atul~K Jain.
\newblock Global greenhouse gas emissions from animal-based foods are twice those of plant-based foods.
\newblock {\em Nature Food}, 2(9):724--732, 2021.

\bibitem{zhong2020associations}
Victor~W Zhong, Linda Van~Horn, Philip Greenland, Mercedes~R Carnethon, Hongyan Ning, John~T Wilkins, Donald~M Lloyd-Jones, and Norrina~B Allen.
\newblock Associations of processed meat, unprocessed red meat, poultry, or fish intake with incident cardiovascular disease and all-cause mortality.
\newblock {\em JAMA internal medicine}, 180(4):503--512, 2020.

\end{thebibliography}

\clearpage

\appendix
\renewcommand{\thesection}{\Alph{section}}

\begin{center}
    \Large
    \SuppI{}
\end{center}

\vspace*{\fill}

\section{Results of On-MFD Analysis}
\xhdr{MFD reduces immediate greenhouse gas emissions (RQ1)}
Table \ref{tab:GHG_emissions_on_MFD} reports the estimated absolute change in GHG emissions associated with Meat-Free Days (MFDs), as captured by the coefficient $\Delta$ in Eq.~1 of the main manuscript. 

To examine the consistency of this effect over time, we also performed a month-stratified analysis. Results are shown in Table~\ref{tab:GHG_emissions_monthly_on_MFD}, where each $\Delta$ coefficient captures the emission reduction associated with MFDs in the corresponding month.

\begin{table}[!ht]
\centering
\begin{tabular}{p{5cm}ccc}
\toprule
\textbf{Variable} &    \textbf{Coefficient} & \textbf{Std.Error} &  \textbf{P-value} \\ 
\midrule
Intercept & 9534.05 & 299.18  & 0.000   \\
$\Delta_{\text{GHG}}$ & -5053.02 & 423.10  & 0.000  \\
\midrule
\textbf{Model Summary:} &  \\
Number of Observations: & 134 \\
Df Residuals: & 132 \\
Pseudo R-squared.: & 0.519 \\
Log-Likelihood: & -1234.8\\
AIC: & 2474\\ \bottomrule
\end{tabular}
\caption{Regression results to capture the variation of GHG emissions in the \emph{on}-MFD setup}
\label{tab:GHG_emissions_on_MFD}
\end{table}

\begin{table}[!ht]
\centering
\begin{tabular}{p{5cm}ccc}
\toprule
\textbf{Variable} &    \textbf{Coefficient} & \textbf{Std.Error} &  \textbf{P-value} \\ 
\midrule
Intercept & 10800.05 & 895.82  & 0.000   \\
$\Delta_{\text{GHG}} \cdot \text{Jan}$ &  -5944.52 &  1368.38   & 0.000  \\
$\Delta_{\text{GHG}} \cdot \text{Feb}$ &  -4845.47 &  1462.86   & 0.000  \\
$\Delta_{\text{GHG}} \cdot \text{Mar}$ &  -4805.78 &  823.20    & 0.001  \\
$\Delta_{\text{GHG}} \cdot \text{Apr}$ &  -5990.00 &  1034.40   & 0.000  \\
$\Delta_{\text{GHG}} \cdot \text{May}$ &  -6439.78 &  895.82    & 0.000  \\
$\Delta_{\text{GHG}} \cdot \text{Jun}$ &  -4720.25 &  895.82    & 0.000  \\
$\Delta_{\text{GHG}} \cdot \text{Jul}$ &  -3158.16 &  927.26    & 0.001  \\
$\Delta_{\text{GHG}} \cdot \text{Aug}$ &  -2800.60 &  996.77    & 0.006  \\
$\Delta_{\text{GHG}} \cdot \text{Sep}$ &  -5935.98 &  1368.38   & 0.000  \\
$\Delta_{\text{GHG}} \cdot \text{Oct}$ & -5968.80 &  1266.88   & 0.000  \\
$\Delta_{\text{GHG}} \cdot \text{Nov}$ & -6210.97 &  1201.87   & 0.000  \\
$\Delta_{\text{GHG}} \cdot \text{Dec}$ & -4354.88 &  1551.60   & 0.000  \\
\midrule 
\textbf{Model Summary:} &  \\
Number of Observations: & 134 \\
Df Residuals: & 110 \\
Pseudo R-squared.: & 0.519 \\
Log-Likelihood: & -1234.8\\
AIC: & 2409\\ \bottomrule
\end{tabular}
\caption{Regression results to capture the monthly variation of GHG emissions in the \emph{on}-MFD setup}
\label{tab:GHG_emissions_monthly_on_MFD}
\end{table}

\newpage

\xhdr{MFD increases off-campus dining (RQ2)} Table \ref{tab:sales_on_MFD_indicator} reports the estimated absolute change in sales associated with Meat-Free Days (MFDs). Table \ref{tab:sales_by_condition} shows the average number of sales on non-MFD vs MFD days.

\begin{table}[!ht]
\centering
\begin{tabular}{p{5cm}ccc}
\toprule
\textbf{Variable} & \textbf{Coefficient} & \textbf{Std. Error} & \textbf{P-value} \\
\midrule
Intercept & 3091.42 & 99.61 & 0.000 \\
 $\Delta_{\text{sales}}$ & -519.93 & 140.87 & 0.000 \\
\midrule
\textbf{Model Summary:} & & & \\
Number of Observations & 134 & & \\
R-squared & 0.094 & & \\
Adjusted R-squared & 0.087 & & \\
AIC & 2179.0 & & \\
\bottomrule
\end{tabular}
\caption{OLS regression estimating the effect of Meat-Free Day (MFD) on sales. $\Delta^{\text{sales}}$ in eq.1 represents the absolute change in meals sold on MFDs vs. non-MFDs.}
\label{tab:sales_on_MFD_indicator}
\end{table}

\begin{table}[!ht]
\centering
\begin{tabular}{p{5cm}ccc}
\toprule
\textbf{Condition} & \textbf{Sales Estimate} & \textbf{Std. Error} & \textbf{P-value} \\
\midrule
Control (non-MFD) & 3091.42 & 99.61 & 0.000 \\
Treatment (MFD) & 2571.49 & 99.61 & 0.000 \\
\midrule
\textbf{Model Summary:} & & & \\
Number of Observations & 134 & & \\
R-squared & 0.094 & & \\
Adjusted R-squared & 0.087 & & \\
AIC & 2179.0 & & \\
\bottomrule
\end{tabular}
\caption{OLS regression estimating the effect of Meat-Free Day (MFD) on sales. Coefficients represent average sales on non-MFDs (control) and MFDs (treatment), respectively.}
\label{tab:sales_by_condition}
\end{table}

\newpage

\xhdr{MFD mantains caloric intake but affects macronutrient mix (RQ3)}
The European Food Safety Authority (EFSA) Panel on Dietetic Products, Nutrition, and Allergies (NDA) provides guidance on the
translation of nutrient based dietary advice into guidance, intended for the European population as a whole, on
the contribution of different foods or food groups to an overall diet that would help to maintain good health
through optimal nutrition (food-based dietary guidelines). The main focus of this Opinion is put on the scientific
process of developing food-based dietary guidelines (FBDG) for the diverse European populations, following a
stepwise approach which  consist of: 1) Identification of diet-health relationships, 2) Identification
of country specific diet-related health problems, 3) Identification of nutrients of public health importance,
4) Identification of foods relevant for FBDG, 5) Identification of food consumption patterns, 6) Testing and
optimising FBDG. 

Based on this approach the FBDG determines the quantities of daily calories and as a consequences of macro- and micro-nutrients for an average individual in order to sustain a healthy and balanced life style.
We report here only macro nutrients as these were the only once to have access to for the analysis in RQ3. 

\begin{itemize}
    \item \textbf{Energy (Calories):} 2000 - 2500 kcal/day
    \item \textbf{Macronutrients:}
        \item \textbf{Proteins:} 0.83 g/kg of body weight/day (approximately 58 g/day for a 70 kg adult)
        \item \textbf{Total Fat:} 20\%-35\% of total energy intake
        \item \textbf{Saturated Fat:} Less than 10\% of total energy intake
        \item \textbf{Monounsaturated Fat:} 10\%-20\% of total energy intake
        \item \textbf{Polyunsaturated Fat:} 6\%-11\% of total energy intake
        \item \textbf{Carbohydrates:} 45\%-60\% of total energy intake
        \item \textbf{Sugars:} Less than 10\% of total energy intake
        \item \textbf{Dietary Fiber:} At least 25 g/day
\end{itemize}

\subsubsection*{Nutrients Analysis}\pdfbookmark[3]{Nutrients Analysis}{NutrientsAnalysis}

Table \ref{tab:macronutrient_interactions} reports the regression outcomes for the absolute changes in grams for the macronutrients analyzed. To provide context, we also calculate the overall percentage change by comparing these shifts to baseline values observed on non-Meat-Free Days (non-MFDs), which are included in the table for reference. This comparison allows us to quantify the relative impact of MFDs on nutrient intake, highlighting both increases and decreases across key dietary components.

\begin{table}[!ht]
\centering
\begin{tabular}{p{6.5cm}ccc}
\toprule
\textbf{Variable} & \textbf{Coefficient} & \textbf{Std. Error} & \textbf{P-value} \\
\midrule
Non-MFD $\times$ Calories & 682.92 & 16.63 & 0.000 \\
$\Delta_{\text{Calories}}$ & -4.02 & 23.52 & 0.864 \\
Non-MFD $\times$ Carbohydrates & 84.34 & 5.33 & 0.000 \\
$\Delta_{\text{Carbohydrates}}$     & 17.38 & 3.49 & 0.000 \\
Non-MFD $\times$ Cholesterol & 4.71 & 0.53 & 0.000 \\
$\Delta_{\text{Cholesterol}}$ & -0.23 & -0.120 & 0.005 \\
Non-MFD $\times$ Fiber & 8.92 & 0.38 & 0.000 \\
$\Delta_{\text{Fiber}}$ & 2.40 & 0.54 & 0.000 \\
Non-MFD $\times$ Monounsaturated Fat & 6.40 & 0.26 & 0.000 \\
$\Delta_{\text{Monounsaturated-Fat}}$ & 0.97 & 0.367 & 0.009 \\
Non-MFD $\times$ Polyunsaturated Fat & 3.60 & 0.229 & 0.000 \\
$\Delta_{\text{Polyunsaturated-Fat}}$ & 1.19 & 0.000 & 0.000 \\
Non-MFD $\times$ Protein & 21.54 & 0.48 & 0.000 \\
$\Delta_{\text{Protein}}$ & -5.94 & 0.678 & 0.000 \\
Non-MFD $\times$ Saturated Fat & 3.59 & 0.18 & 0.000 \\
$\Delta_{\text{Saturated-Fat}}$ & 0.26 & 0.250 & 0.294 \\
Non-MFD $\times$ Sodium & 3.32 & 0.337 & 0.000 \\
$\Delta_{\text{Sodium}}$ & -0.23 & 0.477 & 0.626 \\
Non-MFD $\times$ Sugar & 4.15 & 0.167 & 0.000 \\
$\Delta_{\text{Sugar}}$ & 1.42 & 0.237 & 0.000 \\
\midrule
\textbf{Model Summary:} & & & \\
Number of Observations & 1340 & & \\
R-squared & 0.956 & & \\
Adjusted R-squared & 0.955 & & \\
AIC & 13940.0 & & \\
\bottomrule
\end{tabular}
\caption{OLS regression estimating the association between macronutrients and the outcome variable, stratified by Meat-Free Day (MFD) condition. Each coefficient captures the estimated effect of the corresponding macronutrient on the outcome under non-MFD or MFD conditions.}
\label{tab:macronutrient_interactions_}
\end{table}

\subsubsection*{Consumption of Unhealthy Items}

Table \ref{tab:reg_treatment_unhealthy} reports the regression results capturing the absolute change in the consumption of unhealthy items, categorized into four distinct groups: desserts and sweets, sugary drinks, coffee and hot drinks, and vending machine items. This classification allows us to clearly quantify the impact of Meat-Free Days (MFDs) on the selection of these potentially harmful products, providing insights into behavioral shifts in consumer choices.

\begin{table}[!ht]
\centering
\begin{tabular}{p{5cm}ccc}
\toprule
\textbf{Variable} &    \textbf{Coefficient} & \textbf{Std.Error} &  \textbf{P-value} \\ 
\midrule
Intercept & 37.6145 & 5.319 & 0.000 \\
$\Delta_{1} \cdot \text{Desserts and Sweets}$ & 166.1951 & 7.500 & 0.000 \\
$\Delta_{2} \cdot \text{Pastries and Bakery Items}$ & -15.9734 & 7.642 & 0.037 \\
$\Delta_{3} \cdot \text{Sugary Drinks}$ & 4.3022 & 7.500 & 0.566 \\
$\Delta_{4} \cdot \text{C(treatment):Coffee and Hot Drinks}$ & 0.5796 & 7.959 & 0.942 \\
$\Delta_{5} \cdot \text{C(treatment):Desserts and Sweets}$ & 26.5785 & 7.938 & 0.001 \\
$\Delta_{6} \cdot \text{C(treatment):Pastries and Bakery Items}$ & -2.7604 & 8.072 & 0.733 \\
$\Delta_{7} \cdot \text{C(treatment):Sugary Drinks}$ & 0.2475 & 7.938 & 0.975 \\
\midrule 
\textbf{Model Summary:} &  \\
Number of Observations: & 597 \\
Df Residuals: & 589 \\
Pseudo R-squared.: & 0.733 \\
Log-Likelihood: & -3159.9\\
AIC: & 6336\\ 
\bottomrule
\end{tabular}
\caption{Regression table capturing the variation of purchases of unhealthy and sugary drinks.}
\label{tab:reg_treatment_unhealthy}
\end{table}
\clearpage

\newpage

\section{Results on beyond-MFD Analysis}

\xhdr{MFD triggers rebound in GHG emissions} 
Table \ref{tab:GHG_emissions_on_MFD_rebound} shows the rebound effect in GHG emissions in the beyond-MFD analysis setup.
\begin{table}[!ht]
\centering
\begin{tabular}{p{5cm}ccc}
\toprule
\textbf{Variable} & \textbf{Coefficient} & \textbf{Std. Error} & \textbf{P-value} \\
\midrule
Intercept & 9275.61 & 310.45 & <0.001 \\
$\Delta$ (MFD effect) & 414.20 & 86.30 & <0.001 \\
\midrule
\textbf{Model Summary:} & & & \\
Number of Observations & 112 & & \\
Df Residuals & 110 & & \\
Pseudo R-squared & 0.527 & & \\
Log-Likelihood & -1031.7 & & \\
AIC & 2071.4 & & \\
\bottomrule
\end{tabular}
\caption{Regression results estimating the variation in GHG emissions considering the beyond-MFD setup.
The coefficient $\Delta$ captures the average change in emissions of GHG on a day following MFD compared to another non-MFD occurred at least two days after MFD.}
\label{tab:GHG_emissions_on_MFD_rebound}
\end{table}

\xhdr{MFD triggers rebound in animal-based meals sales (RQ2)}
Table \ref{tab:co2_regr_rebound} summarizes the regression results quantifying the overall change in the percentage of animal-based meals consumed on the day following Meat-Free Days (MFDs), illustrating the rebound effect described in Study 2. Table \ref{tab:co2_regr_rebound_strata} provides the corresponding results stratified by the months included in our analysis.
\begin{table}[!ht]
\centering
\begin{tabular}{p{5cm}ccc}
\toprule
\textbf{Variable} &    \textbf{Coefficient} & \textbf{Std.Error} &  \textbf{P-value} \\ 
\midrule
Intercept & 0.57 & 0.009  & 0.000   \\
$\Delta$ & 0.0345 & 0.013  & 0.008  \\
\midrule
\textbf{Model Summary:} &  \\
Number of Observations: & 123 \\
Df Residuals: & 121 \\
Pseudo R-squared.: & 0.049 \\
Log-Likelihood: & 151.86\\
AIC: & -299.7 \\ \bottomrule
\end{tabular}
\caption{\textbf{Linear Regression for Rebound Effect (Study 2).} $\Delta$ captures the  percentage change in the consumption of animal-based meals in the days following MFDs (see Study 2 ).}
\label{tab:co2_regr_rebound}
\end{table}

\begin{table}[!ht]
\centering
\begin{tabular}{p{5cm}ccc}
\toprule
\textbf{Variable} &    \textbf{Coefficient} & \textbf{Std.Error} &  \textbf{P-value} \\ 
\midrule
Intercept                          &  0.6008  &  0.009  &  0.000 \\
$\Delta_{1} \cdot \text{Jan}$       &  0.0635  &  0.006  &  0.043 \\
$\Delta_{2} \cdot \text{Feb}$       &  0.0712  &  0.009  &  0.084 \\
$\Delta_{3} \cdot \text{Mar}$       &  0.0669  &  0.004  &  0.042 \\
$\Delta_{4} \cdot \text{Apr}$       &  0.0057  &  0.001  &  0.024 \\
$\Delta_{5} \cdot \text{May}$       &  0.0610  &  0.005  &  0.030 \\
$\Delta_{6} \cdot \text{Jun}$       &  0.0387  &  0.007  &  0.070 \\
$\Delta_{7} \cdot \text{Jul}$       &  0.0264  &  0.004  &  0.024 \\
$\Delta_{8} \cdot \text{Aug}$       &  0.0228  &  0.008  &  0.058 \\
$\Delta_{9} \cdot \text{Sep}$       &  0.1808  &  0.004  &  0.019 \\
$\Delta_{10} \cdot \text{Oct}$      &  0.0195  &  0.008  &  0.005 \\
$\Delta_{11} \cdot \text{Nov}$      &  0.0857  &  0.006  &  0.043 \\
$\Delta_{12} \cdot \text{Dec}$      &  0.0467  &  0.009  &  0.031 \\
\midrule 
\textbf{Model Summary:} &  \\
Number of Observations: & 123 \\
Df Residuals: & 99 \\
Adj. R-squared: & 0.119 \\
Log-Likelihood: & 168.88 \\
AIC: & -289.8 \\ \bottomrule
\end{tabular}

\caption{\textbf{Stratified Linear Regression for Rebound Effect (Study 2).} $\Delta$ captures the  percentage change in the consumption of animal-based meals in the days following MFDs per each month .}

\label{tab:co2_regr_rebound_strata}
\end{table}

\clearpage

\xhdr{MFD does not trigger nutritional rebound (RQ3)} Table \ref{tab:nonmfd_macronutrient_interactions} shows the variation in macronutrients in the beyond-MFD analysis setup.

\begin{table}[!ht]
\centering
\begin{tabular}{p{7cm}ccc}
\toprule
\textbf{Variable} & \textbf{Coefficient} & \textbf{Std. Error} & \textbf{P-value} \\
\midrule
Non-MFD$_{+1}$ × Calories & -2.98 & 6.73 & 0.658 \\
Non-MFD$_{+1}$ × Carbs & 15.22 & 7.40 & 0.040 \\
Non-MFD$_{+1}$ × Cholesterol & 1.37 & 6.99 & 0.844 \\
Non-MFD$_{+1}$ × Fiber & 3.12 & 6.86 & 0.650 \\
Non-MFD$_{+1}$ × Monounsaturated Fat & 0.75 & 6.57 & 0.910 \\
Non-MFD$_{+1}$ × Polyunsaturated Fat & 1.04 & 7.01 & 0.881 \\
Non-MFD$_{+1}$ × Protein & -6.59 & 6.80 & 0.332 \\
Non-MFD$_{+1}$ × Saturated Fat & 0.19 & 6.89 & 0.978 \\
Non-MFD$_{+1}$ × Sodium & -1.33 & 7.20 & 0.854 \\
Non-MFD$_{+1}$ × Sugar & 1.20 & 7.09 & 0.866 \\
\midrule
\textbf{Model Summary:} & & & \\
Number of Observations & 1340 & & \\
R-squared & 0.956 & & \\
Adjusted R-squared & 0.955 & & \\
AIC & 13940.0 & & \\
\bottomrule
\end{tabular}
\caption{OLS regression estimates for the interaction between macronutrients. The coefficients captures the variations in macronutriends between non-MFDs occurred one day after the MFD and non-MFDs that occurred at least two days after MFD. This is coherent with the beyond-MFD setup illustrated in Fig.1}
\label{tab:nonmfd_macronutrient_interactions}
\end{table}


\clearpage

\section{Robustness Checks}

\xhdr{Placebo Test}
In this robustness checks we re run again all our analysis of the on-MFD setup. We compared same-week days where the two days compared occured in two consecutive weeks but in absence of the intervention (Meat-Free Day). When conducting this analysis we do not find any statistically signicant differences with regards to any of the variables considered. We provide the results for GHG emissions and sales in  \ref{tab:placebo_emissions_regr} and \ref{tab:placebo_regr_sales} and for the variation of nutrional intake and unhealthy food in table \ref{tab:nutrient_reg}.

\begin{table}[!ht]
\centering
\begin{tabular}{p{5cm}ccc}
\toprule
\textbf{Variable}  &    \textbf{Coefficient} & \textbf{Std.Error} &  \textbf{P-value} \\ 
\midrule
Intercept         & 9487.20 & 529.17  & 0.000   \\
$\Delta_{GHG}$ & 92.32 & 742.83  & 0.901  \\
\midrule
\textbf{Model Summary:} &  \\
Number of Observations: & 536 \\
Df Residuals: & 536 \\
Adj. R-squared: & -0.015 \\
Log-Likelihood: & -631.36 \\
AIC: & 1267 \\
\bottomrule
\end{tabular}
\caption{Placebo Testing on variations of carbon emissions between days of two consecutive weeks that were not MFDs}
\label{tab:placebo_emissions_regr}
\end{table}

\begin{table}[!ht]
\centering
\begin{tabular}{p{5cm}ccc}
\toprule
\textbf{Variable} &    \textbf{Coefficient} & \textbf{Std.Error} &  \textbf{P-value} \\ 
\midrule
Intercept & 2915.98 & 96.48  & 0.000   \\
$\Delta$ & -56.602 & 210.206  & 0.788  \\
\midrule
\textbf{Model Summary:} &  \\
Number of Observations: & 536 \\
Df Residuals: & 530 \\
Pseudo R-squared.: & -0.007 \\
Log-Likelihood: & -1431.6\\
AIC: & 2867 \\ \bottomrule
\end{tabular}
\caption{Placebo Testing on variations of sales between days of two consecutive weeks that were not MFDs}
\label{tab:placebo_regr_sales}
\end{table}

\begin{table}[!ht]
\centering
\begin{tabular}{p{5cm}ccc}
\toprule
\textbf{Variable} &    \textbf{Coefficient} & \textbf{Std.Error} &  \textbf{P-value} \\ 
\midrule
non-MFD Saturated Fat       & 3.4700   & 0.246  & 0.000 \\
Saturated Fat               & 0.2349   & 0.345  & 0.498 \\
non-MFD Monounsaturated Fat & 6.2590   & 0.354  & 0.000 \\
Monounsaturated Fat         & 0.2799   & 0.497  & 0.575 \\
non-MFD Polyunsaturated Fat & 3.6031   & 0.239  & 0.000 \\
Polyunsaturated Fat         & -0.0140  & 0.336  & 0.967 \\
non-MFD Protein             & 21.6681  & 0.677  & 0.000 \\
Protein                     & -0.2488  & 0.950  & 0.794 \\
non-MFD Carbohydrates       & 85.7845  & 3.022  & 0.000 \\
Carbohydrates               & -2.8379  & 4.243  & 0.506 \\
non-MFD Fiber               & 9.1565   & 0.387  & 0.000 \\
Fiber                       & -0.4733  & 0.543  & 0.386 \\
non-MFD Sugar               & 4.1338   & 0.192  & 0.000 \\
Sugar                       & 0.0375   & 0.270  & 0.890 \\
non-MFD Cholesterol         & 4.6990   & 0.729  & 0.000 \\
Cholesterol                 & 0.0221   & 1.024  & 0.983 \\
non-MFD Sodium              & 3.3111   & 0.496  & 0.000 \\
Sodium                      & 0.0253   & 0.697  & 0.971 \\
non-MFD Calories            & 686.0762 & 23.733 & 0.000 \\
Calories                    & -6.2153  & 33.316 & 0.853 \\
\midrule 
\textbf{Model Summary:} &  \\
Number of Observations: & 5360 \\
Df Residuals: & 5340 \\
Average R-squared: & 0.0038 \\
Average Log-Likelihood: & -443.19 \\
Average AIC: & 803.87 \\
\bottomrule
\end{tabular}
\caption{Placebo testing on variations of key macronutrients and calories when comparing non-MFDs of consecutive weeks}
\label{tab:nutrient_reg}
\end{table}

\clearpage

\xhdr{Comparison with subsequent week}. 
We re-conducted our analysis using the same experimental setup of the on-and beyond-MFD. Specifically, for the on-MFD setup,   we  matchMeat-Free Days (MFDs) with the non-MFDs of the \emph{following} week, rather than the \emph{previous} week. We do the same for the beyond-MFD setup. We match the non-MFD occurred one day after MFD with a non-MFD that occurred at least two days after the MFD of the following week. analysi reduction of the carbon footprint. We reconducted our analysis considering this setup, finding that our results remain robust again this variation.

\begin{table}[!ht]
\centering
\begin{tabular}{p{5cm}ccc}
\toprule
\textbf{Variable} & \textbf{Coefficient} & \textbf{Std. Error} & \textbf{P-value} \\ 
\midrule
Intercept & 9478.62 & 312.05 & 0.000 \\
$\Delta$ & -4987.41 & 415.27 & 0.000 \\
\midrule
\textbf{Model Summary:} & & & \\
Number of Observations & 130 & & \\
Df Residuals & 128 & & \\
Pseudo R-squared & 0.514 & & \\
Log-Likelihood & -1196.2 & & \\
AIC & 2398.3 & & \\
\bottomrule
\end{tabular}
\caption{Regression results estimating the variation in GHG emissions in the \emph{on}-MFD setup. The coefficient $\Delta$ reflects the average reduction in emissions on Meat-Free Days.}
\label{tab:GHG_emissions_on_MFD_variant}
\end{table}

\begin{table}[!ht]
\centering
\begin{tabular}{p{5cm}ccc}
\toprule
\textbf{Variable} & \textbf{Coefficient} & \textbf{Std. Error} & \textbf{P-value} \\
\midrule
Intercept & 3110.35 & 101.42 & 0.000\\
$\Delta^{\text{sales}}$ & -504.76 & 138.21 & 0.003 \\
\midrule
\textbf{Model Summary:} & & & \\
Number of Observations & 134 & & \\
R-squared & 0.091 & & \\
Adjusted R-squared & 0.084 & & \\
AIC & 2181.2 & & \\
\bottomrule
\end{tabular}
\caption{OLS regression estimating the effect of Meat-Free Day (MFD) on sales. $\Delta^{\text{sales}}$ in Eq.~1 represents the absolute change in meals sold on MFDs vs. non-MFDs.}
\label{tab:sales_on_MFD_indicator_var}
\end{table}

\begin{table}[!ht]
\centering
\begin{tabular}{p{6.5cm}ccc}
\toprule
\textbf{Variable} & \textbf{Coefficient} & \textbf{Std. Error} & \textbf{P-value} \\
\midrule
Non-MFD $\times$ Calories & 681.17 & 16.85 & 0.000 \\
MFD $\times$ Calories & -3.88 & 23.67 & 0.866 \\
Non-MFD $\times$ Carbohydrates & 83.72 & 5.51 & 0.000 \\
MFD $\times$ Carbohydrates     & 17.12 & 3.58 & 0.000 \\
Non-MFD $\times$ Cholesterol & 4.67 & 0.55 & 0.000 \\
MFD $\times$ Cholesterol & -0.091 & 0.121 & 0.907 \\
Non-MFD $\times$ Fiber & 8.86 & 0.39 & 0.000 \\
MFD $\times$ Fiber & 2.34 & 0.56 & 0.000 \\
Non-MFD $\times$ Monounsaturated Fat & 6.37 & 0.27 & 0.000 \\
MFD $\times$ Monounsaturated Fat & 0.94 & 0.371 & 0.010 \\
Non-MFD $\times$ Polyunsaturated Fat & 3.58 & 0.235 & 0.000 \\
MFD $\times$ Polyunsaturated Fat & 1.15 & 0.001 & 0.000 \\
Non-MFD $\times$ Protein & 21.33 & 0.49 & 0.000 \\
MFD $\times$ Protein & -5.81 & 0.689 & 0.000 \\
Non-MFD $\times$ Saturated Fat & 3.57 & 0.19 & 0.000 \\
MFD $\times$ Saturated Fat & 0.24 & 0.252 & 0.301 \\
Non-MFD $\times$ Sodium & 3.29 & 0.342 & 0.000 \\
MFD $\times$ Sodium & -0.25 & 0.481 & 0.633 \\
Non-MFD $\times$ Sugar & 4.12 & 0.172 & 0.000 \\
MFD $\times$ Sugar & 1.39 & 0.239 & 0.000 \\
\midrule
\textbf{Model Summary:} & & & \\
Number of Observations & 1340 & & \\
R-squared & 0.955 & & \\
Adjusted R-squared & 0.954 & & \\
AIC & 13942.1 & & \\
\bottomrule
\end{tabular}
\caption{OLS regression estimating the association between macronutrients and the outcome variable, stratified by Meat-Free Day (MFD) condition. Each coefficient captures the estimated effect of the corresponding macronutrient on the outcome under non-MFD or MFD conditions.}
\label{tab:macronutrient_interactions}
\end{table}

\begin{table}[!ht]
\centering
\begin{tabular}{p{5cm}ccc}
\toprule
\textbf{Variable} & \textbf{Coefficient} & \textbf{Std. Error} & \textbf{P-value} \\
\midrule
Intercept & 9321.78 & 308.67 & 0.000 \\
$\Delta$ (MFD effect) & 407.56 & 85.14 & 0.000 \\
\midrule
\textbf{Model Summary:} & & & \\
Number of Observations & 112 & & \\
Df Residuals & 110 & & \\
Pseudo R-squared & 0.522 & & \\
Log-Likelihood & -1035.8 & & \\
AIC & 2083.6 & & \\
\bottomrule
\end{tabular}
\caption{Regression results estimating the variation in GHG emissions in the beyond-MFD setup.
The coefficient $\Delta$ captures the average change in GHG emissions on a day following an MFD, compared to a non-MFD occurring at least two days after the last MFD.}
\label{tab:GHG_emissions_on_MFD_rebound_}
\end{table}

\begin{table}[!ht]
\centering
\begin{tabular}{p{5cm}ccc}
\toprule
\textbf{Variable} & \textbf{Coefficient} & \textbf{Std. Error} & \textbf{P-value} \\ 
\midrule
Intercept & 0.571 & 0.0091 & 0.000 \\
$\Delta$ & 0.0321 & 0.0126 & 0.012 \\
\midrule
\textbf{Model Summary:} & & & \\
Number of Observations: & 123 & & \\
Df Residuals: & 121 & & \\
Pseudo R-squared: & 0.046 & & \\
Log-Likelihood: & 150.72 & & \\
AIC: & -297.4 & & \\
\bottomrule
\end{tabular}
\caption{\textbf{Linear Regression for Rebound Effect (Study 2).} $\Delta$ captures the percentage change in the consumption of animal-based meals on the day following MFDs (see Study 2).}
\label{tab:co2_regr_rebound_}
\end{table}

\clearpage

\clearpage

\section{Further Specification for the Dataset}

We analyzed all meals offered across the 12 cafeterias during the 134-day study period, which included 67 Meat-Free Days (MFDs) and 67 non-Meat-Free Days (non-MFDs). On non-MFDs, 51.6\% of the meals served on campus were vegetarian, while, as expected, 100\% of meals were vegetarian on MFDs. Table \ref{tab:menu_comparison} presents the average number of vegetarian meals served on both MFDs and non-MFDs.

To explore whether this increase in vegetarian meal production on MFDs affected ingredient usage, we analyzed the detailed ingredient data for each meal. We found an 85\% overlap in vegetarian ingredients between MFDs and non-MFDs, with the remaining 15\% difference stemming from a greater variety of ingredients used on MFDs. For instance, while only four types of rice were used on non-MFDs, eight varieties appeared on MFDs, a pattern similarly observed with legumes, mozzarella, and tomatoes.

Despite this variability, the most frequently used ingredients remained consistent across both MFDs and non-MFDs. Fig.  6 shows the top 50 ingredients used on both types of days, ranked by their relative frequency of use.

\begin{table}[htbp]
\centering
\begin{tabular}{p{5cm}ccc}
\toprule
\textbf{Restaurant} & \textbf{MFDs} & \textbf{non-MFDs} \\
\midrule
Alpine & 4.545455 & 2.050847 \\ 
Arcadie & 9.283582 & 10.453125 \\ 
Epicure & 2.384615 & 2.714286 \\ 
Esplanade & 6.255319 & 4.257143 \\ 
Giacometti & 2.420000 & 1.434783 \\ 
Ginko & 1.243902 & 1.000000 \\ 
Hopper & 4.950820 & 3.679245 \\ 
Le Native & 3.578947 & 3.774194 \\ 
Niki & 3.342857 & 2.200000 \\ 
Ornithorynque & 8.014925 & 4.681818 \\ 
Piano & 10.274194 & 4.983871 \\ 
Puur & 6.017241 & 4.035088 \\ 
Zaha & 2.695652 & 1.916667 \\ 
\bottomrule
\end{tabular}
\caption{Comparison of vegetarian meal percentages served on non-MFDs (second column) and MFDs (third column) across restaurants (first column). Meal averages vary as restaurants don't serve the same number of meals daily}
\label{tab:menu_comparison}
\end{table}

\begin{figure*}[t]
    \centering
    \includegraphics[width=1.0\textwidth]{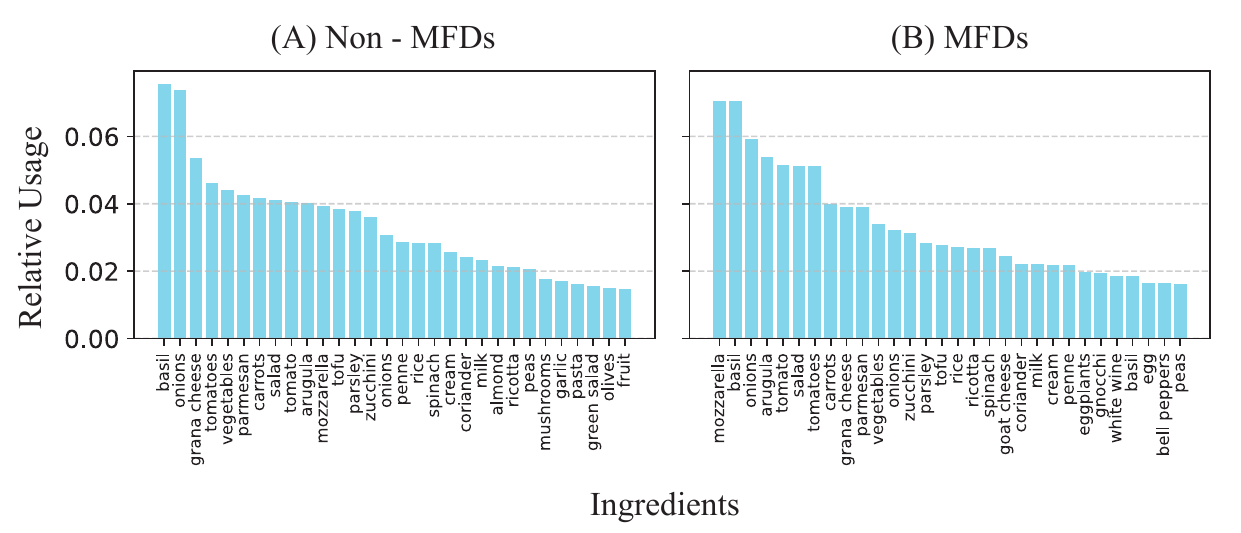}
    \caption{
    \textbf{Comparison of vegeterian ingredients usage in non-MFDs and MFDs.} The left panel (A) shows the relative frequency (y-axis) of the top 30 ingredients used for the meal preparation on non-Meat Free days (Non-MFDs). Similarly, the right panel (B) shows the relative frequencies for MFDs. We note a high overlap between the ingredients used in non-MFDs and MFDs.
    }

    \label{ig:ingredients_rank}
    \vspace{-1mm}

\end{figure*}

\xhdr{Alternative Dining Options} To underscore the potential impact of off-campus dining on the intervention’s effectiveness, we provide a detailed map of the EPFL campus that illustrates the proximity of nearby alternative dining options. A fast-food restaurant and a grocery store are located just 51 and 127 meters, respectively, from the nearest campus building, with the farthest buildings being 482 and 558 meters away, corresponding to a maximum walking distance of approximately seven minutes. As most of the campus community is concentrated around the central buildings, these alternatives are typically just a 3-4 minute walk, making off-campus dining a convenient and viable option.

\begin{figure*}[t]
    \centering
    \includegraphics[width=1.0\textwidth]{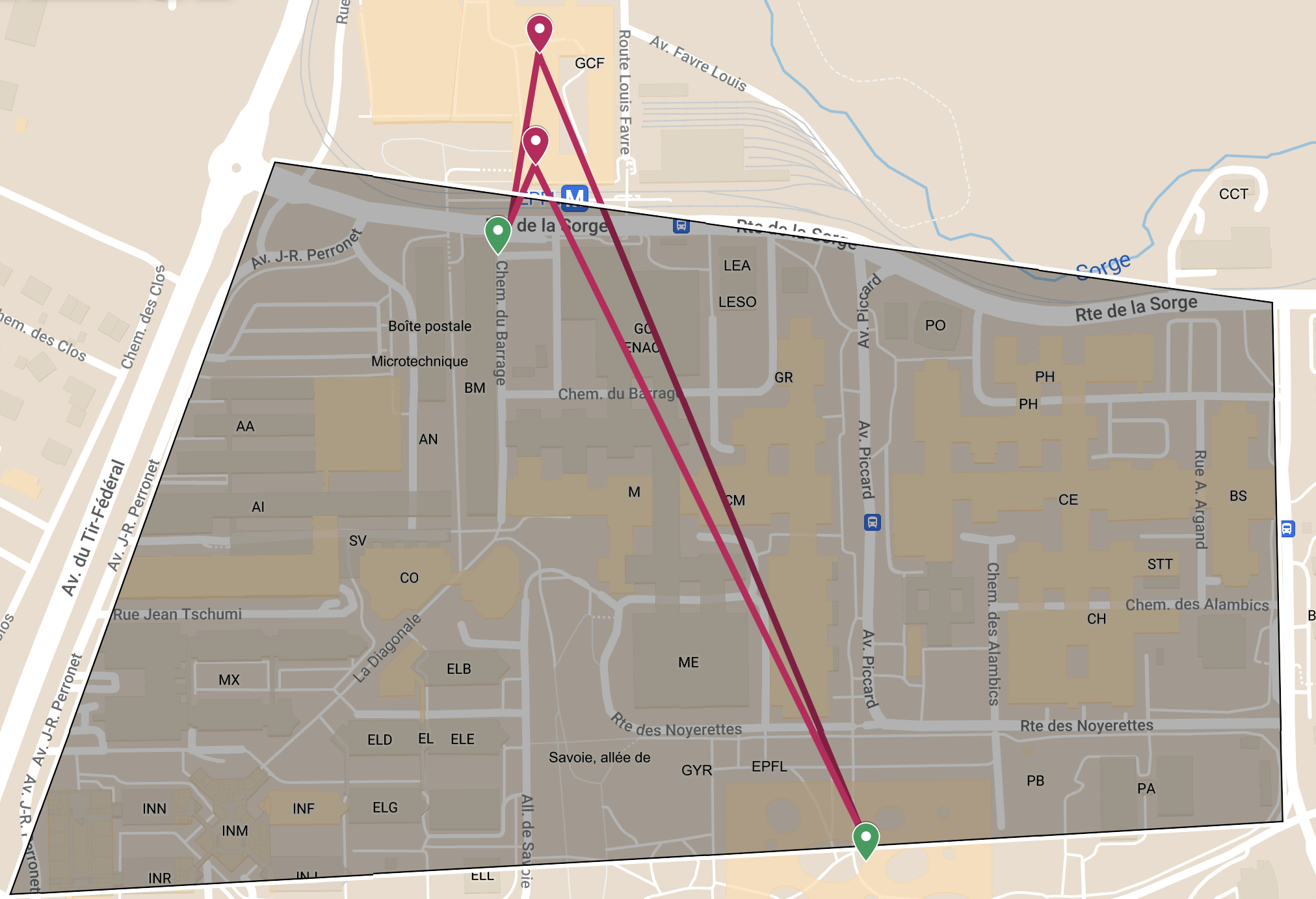}
    \caption{
    \textbf{Campus Map of The EPFL.} In green we indicate respectively the closest and farthest away buildings from the campus entrance at EPFL. In red we indicate the fast-food restaurant (the closest to the campus entrance) and the grocery shop.
    }

    \label{fig:map}
    \vspace{-1mm}

\end{figure*}

\end{document}